\documentclass[11pt,nofootinbib,preprint]{revtex4}
\usepackage{graphicx}
\usepackage{amsmath}
\usepackage{amsfonts}
\usepackage{amssymb}
\usepackage{color}

\newcommand{\omits}[1]{}

\def\bc{\begin{center}}
\def\nno{\nonumber}
\def\ec{\end{center}}

\def\be{\begin{eqnarray}}
\def\ee{\end{eqnarray}}

\begin{document}


\title{Pair production from Reissner-Nordstr{\"o}m-anti-de Sitter black holes}

\author{Jun Zhang$^1$} \email{zhangj626@mail2.sysu.edu.cn}
\author{Yi-Yu Lin$^1$} \email{linyy27@mail2.sysu.edu.cn}
\author{Hao-Chun Liang$^1$} \email{lianghch3@mail2.sysu.edu.cn}
\author{Ke-Jia Chi$^1$} \email{chikj@mail2.sysu.edu.cn}
\author{Chiang-Mei Chen$^2$} \email{cmchen@phy.ncu.edu.tw}
\author{Sang Pyo Kim$^3$} \email{sangkim@kunsan.ac.kr}
\author{Jia-Rui Sun$^1$} \email{sunjiarui@sysu.edu.cn}

\affiliation{${}^1$School of Physics and Astronomy, Sun Yat-Sen University, Guangzhou 510275, China}
\affiliation{${}^2$Department of Physics, and Center for High Energy and High Field Physics (CHiP), National Central University, Chungli 32001, Taiwan}
\affiliation{${}^3$Department of Physics, Kunsan National University, Kunsan 54150, Korea}



\begin{abstract}
We study the pair production of charged scalar particles from the five-dimensional near extremal Reissner-Nordstr\"om-Anti de Sitter (RN-AdS$_5$) black hole. The pair production rate and the absorption cross section ratio in the full spacetime are obtained and are shown to have proportional relation with their counterparts in the near horizon region. In addition, the holographic descriptions of the pair production both in the IR CFT in the near horizon region and the UV CFT at the asymptotic spatial boundary of the RN-AdS$_5$ black hole are analyzed in the AdS$_2$/CFT$_1$ and AdS$_5$/CFT$_4$ correspondences, respectively. This work gives a complete description of scalar pair production in the near extremal RN-AdS$_5$ black hole.
\end{abstract}

\pacs{04.62.+v, 04.70.Dy, 12.20.-m}

\maketitle
\tableofcontents

\section{Introduction}
The Schwinger pair production of charged particles is an important QED phenomenon which is related to the vacuum instability and persistence in the presence of strong external electromagnetic field~\cite{Schwinger:1951nm}. Another important spontaneous pair production phenomenon is the Hawking radiation from black holes, which can be viewed as a tunneling process through the black hole horizon~\cite{Parikh:1999mf}. A charged black hole thus provides a natural lab in which both the Schwinger pair production and the Hawking radiation can occur and mix with each other. Usually the equation of motions (EoMs) of quantum fields in a general black hole background is difficult to be solved analytically in the full spacetime. However, when the symmetry of a spacetime geometry is enhanced under some conditions, the problem becomes manageable; for this reason, in a series of recent studies, the spontaneous pair production of charged particles has been systematically studied in near extremal charged black holes, including the RN black hole~\cite{Chen:2012yn, Chen:2014yfa, Chen:2020mqs} and the Kerr-Newman (KN) black hole~\cite{Chen:2016caa, Chen:2017mnm}, in which the near horizon geometry is enhanced into the AdS$_2$ or warped AdS$_3$ in the near extremal limit. Thanks to the enhanced near horizon symmetry, the explicit forms of the pair production rate and other 2-point correlation functions have been obtained and their holographic descriptions have been found based on the RN/CFT~\cite{Chen:2009ht, Chen:2010bsa, Chen:2010as, Chen:2010yu, Chen:2012ps, Chen:2012pt} and KN/CFTs dualities~\cite{Chen:2010ywa, Hartman:2008pb, Hartman:2009nz}. In addition to charged black hole backgrounds, the pair production has also been investigated in a pure AdS or dS spacetime, see, e.g.,~\cite{Garriga:1994bm, Pioline:2005pf, Kim:2008xv, Cai:2014qba}, whereas in the absence of gravitational field the pure Schwinger effect has been efficiently analyzed by using the phase-integral method~\cite{Kim:2000un, Kim:2007pm, Dumlu:2010ua, Kim:2019yts}.

However, previous studies mainly focus on analyzing the spontaneous pair production in the near horizon region of black holes in an asymptotically flat spacetime. A charged black hole in AdS spacetime has an additional AdS symmetry at the asymptotical boundary. From the holographic point of view, the CFT description of pair production has been revealed only in the near horizon region in terms of the AdS$_2$/CFT$_1$ (or warped AdS$_3$/CFT$_2$). Although particle pairs produced in the near horizon region of black hole indeed give important contributions to those in the full spacetime, understanding of the whole picture is still lacking. In the present paper, we extend the study of pair production to the full near extremal RN-AdS$_5$ black hole background, which possesses an AdS$_5$ geometry at the asymptotic spatial boundary as well as an AdS$_2$ structure in the near horizon region. It is shown that the radial equation of the charged scalar field propagating in this spacetime can be transformed into a Heun-like differential equation and thus be solved by matching its solutions in the near and far spacetime regions. Consequently, analytical forms of the full solutions for the pair production rate, the absorption cross section ratio as well as the retarded Green's functions are obtained and they are shown to have elegant relations with their counterparts calculated in the near horizon region.

The near extremal RN-AdS$_5$ black hole is also a very useful background to study holographic dualities. As the near horizon AdS$_2$ (or warped AdS$_3$) spacetime is dual to a 1D CFT (or chiral CFT$_2$), while the asymptotical AdS$_5$ spacetime is dual to another 4D CFT, the former is called the IR CFT, while the latter is called the UV CFT, and they are connected with each other via the holographic renormalization group (RG) flow along the radial direction~\cite{Balasubramanian:1999, Heemskerk:2010hk, Faulkner:2010jy}. For example, it has been shown that the near extremal RN-AdS black hole acts as a holographic model to describe typical properties of the (non)Fermi liquid at quantum critical point~\cite{Faulkner:2009wj, Iqbal:2009fd, Liu:2009dm, Faulkner:2011tm}. It is thus natural and interesting to find holographic descriptions of pair production in the RN-AdS$_5$ black hole  both in the IR CFT$_1$ in the near horizon region and the UV CFT$_4$ at the asymptotical AdS$_5$ boundary. We show that the picture in the IR CFT$_1$ is very similar to those in the near extremal RN and KN black holes, the pair production rate and the absorption cross section ratio calculated from the AdS$_2$ spacetime can match with those from the dual IR CFT. As for the UV 4D CFT, although a direct comparison of calculations between the bulk and the boundary in terms of the AdS$_5$/CFT$_4$ is not made due to the lack of informations on the dual finite temperature CFT$_4$ side. However, from the bulk gravity side, the condition for pair production in the full near extremal RN-AdS$_5$ spacetime is the violation of the Breitenlohner-Freedman (BF) bound~\cite{Breitenlohner:1982jf, Breitenlohner:1982bm} in AdS$_5$ spacetime. This, on the dual 4D CFT side, corresponds to a complex conformal weight for the scalar operator dual to the bulk charged scalar field, which indeed indicates instabilities for the scalar operator on the boundary and is consistent with the situation in the IR CFT. Furthermore, we find an interesting relation between the full pair production rate and the absorption cross section ratio via changing the roles of sources and operators at the same time both in the IR and the UV CFTs.

The rest of the paper is organized as follows. In sec.~\ref{sec:bgd} we give a brief review of the bulk theory and consider the near horizon geometry of the RN-AdS$_{d+1}$ black hole and the EoMs of the probe charged scalar field. In sec.~\ref{section:Pair production in the inner AdS$_2$}, the spontaneous pair production in the near horizon region of the near extremal RN-AdS$_{d+1}$ black holes is discussed and the 2-point functions of the charged scalar field, such as the retarded Green's function, and the pair production rate and absorption cross section ratio  are calculated. In sec.~\ref{sec:AdS5 pair production}, the full analytical solution for the radial equation of the charged scalar field in RN-AdS$_5$ black holes is obtained by applying the matching technique. Consequently, the full analytical forms of the pair production rate, the absorption cross section ratio, as well as the retarded Green's function are found, and the connection with their counterparts in the near horizon region of the black hole is discussed. Then in sec.~\ref{sec:cft descriptions}, the dual CFTs descriptions of the spontaneous pair production are both analyzed in terms of AdS$_2$/CFT$_1$ correspondence in the IR region and the AdS$_5$/CFT$_4$ correspondence in the UV region, and their connections are also revealed. Finally, the conclusion and physical implications are given in sec.~\ref{conclusion}.

\section{The bulk theory}\label{sec:bgd}
\subsection{The RN-AdS$_{d+1}$ black hole}

The $d+1$ dimensional Einstein-Maxwell theory has the action (in units of $c = \hbar = 1$) as
\begin{equation}
I = \int d^{d+1}x \sqrt{-g} \left[ \frac{1}{16 \pi G_{d+1}} \left( R + \frac{d(d-1)}{L^2} \right) - \frac1{g_{\mathrm s}^2} F_{\mu\nu} F^{\mu\nu} \right],
\end{equation}
where $L$ is the curvature radius of the asymptotically AdS$_{d+1}$ spacetime and $g_{\mathrm s}$ is the dimensionless coupling constant of the $U(1)$ gauge field. The dynamical equations
\begin{eqnarray}
R_{\mu\nu} - \frac12 g_{\mu\nu} R - \frac{d(d-1)}{2L^2} g_{\mu\nu} &=& \frac{8 \pi G_{d+1}}{g_{\mathrm s}^2} \left( 4 F_{\mu\lambda} F_{\nu}{}^{\lambda} - g_{\mu\nu} F_{\alpha\beta} F^{\alpha\beta} \right),
\nonumber\\
\partial_\mu \left( \sqrt{-g} F^{\mu\nu} \right) &=& 0,
\end{eqnarray}
admit the Reissner-Nordstr{\"o}m-Anti de Sitter (RN-AdS$_{d+1}$) black brane (or the planar black hole) solution~\cite{Xu:1988ju}
\begin{eqnarray} \label{RNAdS}
ds^2 &=& \frac{L^2}{r^2 f(r)} dr^2 + \frac{r^2}{L^2} \left( -f(r) dt^2 + dx_i^2 \right),
\nonumber\\
A &=& \mu \left( 1 - \frac{r_{\mathrm o}^{d-2}}{r^{d-2}} \right) dt,
\end{eqnarray}
with
\begin{equation}
f(r) = 1 - \frac{G_{d+1} L^2 M}{r^d} + \frac{G_{d+1} L^2 Q^2}{r^{2d-2}}, \qquad {\rm and} \qquad \mu = \sqrt{\frac{d-1}{2(d-2)}} \frac{g_{\mathrm s} Q}{r_{\mathrm o}^{d-2}},
\end{equation}
where $r_{\mathrm o}$ is the radius of the outer horizon ($f(r_{\mathrm o}) = 0$), $\mu$ is the chemical potential with dimension $[\mu] = \mathrm{length}^{-(d-1)/2}$, $M$ is the mass and $Q$ is the charge of the black brane. We may find an explicit expression of $r_{\mathrm o}$ for $d = 4$ from a solution of a cubic equation, which is complicated, but $r_\mathrm{o}$ has a general expression for the extremal case, i.e. $r_*$ in \ref{nhneg}. The condition $f(r_{\mathrm o}) = 0$ gives $M = \frac{r_{\mathrm o}^d}{G_{d+1} L^2} + \frac{Q^2}{r_{\mathrm o}^{d-2}}$ (which is the Smarr-like relation related to the first law of thermodynamics of the black brane) and the temperature $T$ and ``surface'' entropy density $s$ of the black brane are respectively
\begin{equation}
T = \frac{d \, r_{\mathrm o}}{4\pi L^2} \left( 1 - \frac{d-2}{d} \frac{G_{d+1} L^2 Q^2}{r_{\mathrm o}^{2d-2}} \right), \qquad {\rm and} \qquad s = \frac{1}{4G_{d+1}} \left( \frac{r_{\mathrm o}}{L} \right)^{d-1}.
\end{equation}
Besides, the first law of thermodynamics of the dual boundary $d$-dimensional quantum field is
\begin{equation}
\delta \epsilon = T \delta s + \mu \delta\rho_{\mathrm c},
\end{equation}
where the ``surface'' energy and charge densities are respectively
\begin{equation}
\epsilon = \frac{d - 1}{16 \pi L^{d-1}} M, \qquad {\rm and} \qquad \rho_{\mathrm c} = \frac{\sqrt{2(d-1)(d-2)}}{8 \pi g_{\mathrm s} L^{d-1}} Q.
\end{equation}
Then it is straightforward to check the Euler relation
\begin{equation}
\left( \frac{d}{d-1} \right) \epsilon = \epsilon + p = T s + \mu \rho_{\mathrm c},
\end{equation}
where the pressure is $p = \frac{\epsilon}{d-1}$, which shows that the dual $d$-dimensional quantum field theory on the asymptotic boundary is conformal as expected.

\subsection{Near-horizon near-extremal geometry}\label{nhneg}
To make the following analysis convenient, let us introduce the length scale $r_*^{2d-2} \equiv \frac{d-2}{d} G_{d+1} L^2 Q^2$, then the temperature is rewritten as
\begin{equation}
T = \frac{d r_{\mathrm o}}{4\pi L^2} \left( 1 - \frac{r_*^{2d-2}}{r_{\mathrm o}^{2d-2}} \right).
\end{equation}
Note that $r_*$ may be treated as the ``effective'' radius of the inner black hole horizon though $f(r_*) \neq 0$ in general and $r_* < r_{\mathrm o}$. The extremal condition for a degenerate horizon at $r_\mathrm{o} = r_*$ is $M = M_0 \equiv \frac{2 (d-1)}{d-2} \frac{r_*^d}{G_{d+1} L^2}$. The near extremal limit of near horizon is obtained by taking the limit $\varepsilon \to 0$ of the transformations
\begin{equation}\label{scaling}
M - M_0 = \frac{d (d-1) r_*^{d-2}}{G_{d+1} L^2} \varepsilon^2 \rho_\mathrm{o}^2, \qquad r_{\mathrm o} - r_* = \varepsilon \rho_{\mathrm o}, \qquad r - r_{\mathrm o} = \varepsilon (\rho - \rho_0), \qquad t = \frac{\tau}{\varepsilon},
\end{equation}
where in general $\rho_{\mathrm o}$ is finite and $\rho \in [\rho_0, \infty)$.

Expanding $f(r)$ around $r = r_{\mathrm o}$,
\begin{equation} \label{fr}
f(r) = f(r_{\mathrm o}) + f'(r_{\mathrm o}) (r - r_{\mathrm o}) + \frac{f''(r_{\mathrm o})}{2} (r - r_{\mathrm o})^2 + \mathcal{O}(\varepsilon^3),
\end{equation}
we have
\begin{eqnarray} \label{fr1}
f'(r_{\mathrm o}) &=& \frac{4\pi L^2}{r_{\mathrm o}^2} T \simeq \frac{2 d (d-1) \rho_0}{r_{\mathrm o}^2}\varepsilon,
\nonumber\\
f''(r_{\mathrm o}) &=& \frac{2d(d-1)}{r_{\mathrm o}^2}-\frac{(3d-1)4\pi L^2 T}{r_{\mathrm o}^3} \simeq \frac{2 d (d-1)}{r_{\mathrm o}^2},
\end{eqnarray}
where the temperature is expanded in powers of $\varepsilon$ as $T = \frac{d (d-1)}{2 \pi L^2} \rho_{\mathrm o} \varepsilon + \mathcal{O}(\varepsilon^2)$. Finally, with
\begin{equation}
f(r) \simeq \frac{d (d-1)}{r_{\mathrm o}^2} (\rho^2 - \rho_{\mathrm o}^2) \varepsilon^2,
\end{equation}
the near horizon geometry is given by
\begin{eqnarray}\label{nhne1}
ds^2 &=& - \frac{\rho^2 - \rho_{\mathrm o}^2}{\ell^2} d\tau^2 + \frac{\ell^2 d\rho^2}{\rho^2 - \rho_{\mathrm o}^2} + \frac{r_{\mathrm o}^2}{L^2} dx_i^2,
\nonumber\\
A &=& \frac{(d-2) \mu}{r_{\mathrm o}} (\rho - \rho_{\mathrm o}) d\tau,
\end{eqnarray}
where $\ell^2 \equiv \frac{L^2}{d (d - 1)}$ is defined as the square of curvature radius of the effective AdS$_2$ geometry. The limit $\rho_{\mathrm o} \to 0$ yields the extremal limit.

The solution in eq.(\ref{nhne1}) can also be written in the Poincar\'{e} coordinates in terms of $\xi = \ell^2/\rho$, ($|\xi| \leq \xi_{\mathrm o} = \ell^2/\rho_\mathrm{o}$),
\begin{eqnarray} \label{NRNAdS}
ds^2 &=& \frac{\ell^2}{\xi^2} \left( - \left( 1 - \frac{\xi^2}{\xi_{\mathrm o}^2} \right) d\tau^2 + \frac{d\xi^2}{1-\frac{\xi^2}{\xi_{\mathrm o}^2}} \right) + \frac{r_{\mathrm o}^2}{L^2} dx_i^2,
\nno\\
A &=& \frac{(d-2) \mu \ell^2}{r_{\mathrm o}} \left( \frac{1}{\xi} - \frac{1}{\xi_{\mathrm o}} \right) d\tau.
\end{eqnarray}
The above geometry is a black brane with both local and asymptotical topology  AdS$_2 \times \mathbf{R}^{d-1}$ (the AdS$_2$ has the $SL(2,R)_R$ symmetry). The horizons of the new black brane are located at $\xi = \pm \xi_{\mathrm o}$ and its temperature is $T_{\mathrm n} = \frac{1}{2 \pi \xi_{\mathrm o}}$. Note that if we adopt the new coordinates $z \equiv \xi/\xi_{\mathrm o}$ with $|z| \leq 1$ and $\eta = \tau/\xi_{\mathrm o}$, then the metric becomes
\begin{equation} \label{NRNAdS1}
ds^2 = \frac{\ell^2}{z^2} \left( - (1 - z^2) d\eta^2 + \frac{dz^2}{1-z^2} \right) + \frac{r_{\mathrm o}^2}{L^2} dx_i^2,
\end{equation}
and the temperature associated with the inverse period of $\eta$ is normalized to $\tilde{T}_{\mathrm n} = \frac{1}{2\pi}$.

\subsection{Charged scalar field probe}
The action of a bulk probe charged scalar field $\Phi$ with mass $m$ and charge $q$ is
\begin{equation} \label{action}
S = \int {{d^{d + 1}}} x\sqrt { - g} \left( { - \frac{1}{2}{D_\alpha^* }{\Phi ^*}{D^\alpha }\Phi  - \frac{1}{2}{m^2}{\Phi ^*}\Phi } \right),
\end{equation}
where $D_{\alpha} \equiv \nabla_{\alpha} - i q A_{\alpha}$ with $\nabla_\alpha$ being the covariant derivative in curved spacetime. The corresponding Klein-Gordon (KG) equation is
\begin{equation} \label{eom1}
(\nabla_\alpha - i q A_\alpha) (\nabla^\alpha - i q A^\alpha) \Phi = m^2 \Phi.
\end{equation}
Besides, the radial flux of the probe field is
\begin{equation} \label{flux}
\mathcal{F} = i\sqrt { - g} {g^{rr}}(\Phi D_r^*{\Phi ^*} - {\Phi ^*}{D_r}\Phi ).
\end{equation}

In the RN-AdS$_{d+1}$ background~(\ref{RNAdS}), assuming $\Phi(t, \vec{x}, r) = \phi(r) \mathrm{e}^{-i \omega t + i \vec{k} \cdot \vec{x}}$, the KG equation~(\ref{eom1}) has the radial equation
\begin{equation} \label{eomtr}
\left(\frac{L}{r}\right)^{d-1} \partial_r \left( \frac{r^{d+1}}{L^{d+1}} f(r) \partial_r \right) \phi(r) + \left(\frac{L^2 (\omega + q A_t)^2}{r^2 f(r)} - m^2 - \frac{L^2}{r^2} \vec{k}^2 \right) \phi(r) = 0.
\end{equation}
The solutions to eq.~(\ref{eomtr}) cannot be directly found in terms of special functions in the full spacetime region. In the following, we will solve it in different regions respectively and match these solutions to obtain the full solution.

\section{Pair production in the inner AdS$_2$}\label{section:Pair production in the inner AdS$_2$}
\subsection{Near-horizon solutions}
Firstly, we analyze the near horizon, near extreme region~(\ref{NRNAdS}) and solve the KG equation~(\ref{eom1}) by expanding the scalar field as
\begin{equation} \label{ansatz1}
\Phi(\tau, \vec{x}, \xi) = \phi(\xi) \mathrm{e}^{-i w \tau + i \vec{k} \cdot \vec{x}}.
\end{equation}
Then the KG equation reduces to\footnote{Note that from eq.~(\ref{scaling}), one has $\omega=\varepsilon w$, which may indicate that if one considered the finite frequency probe field in the $\tau$, $\xi$ coordinates, the probe field in the original $t$, $r$ coordinates is automatically driven into the low frequency region.}
\begin{equation} \label{eom2}
\xi^2 \left( 1 - \frac{\xi^2}{\xi_{\mathrm o}^2} \right) \phi''(\xi) - \frac{2\xi^3}{\xi_{\mathrm o}^2} \phi'(\xi) + \xi^2 \frac{(w + q A_\tau)^2}{1 - \frac{\xi^2}{\xi_{\mathrm o}^2}} \phi(\xi) = m_\mathrm{eff}^2 \ell^2 \phi(\xi),
\end{equation}
where the effective mass square is defined as $m_\mathrm{eff}^2 = m^2 + \frac{L^2 \vec{k}^2}{r_{\mathrm o}^2}$. Or, the KG equation can be expressed in the $z$ coordinate as
\begin{equation} \label{eom3}
z^2 (1 - z^2) \phi''(z) - 2 z^3 \phi'(z) + \frac{z^2}{1 - z^2} \left[ \left( w \xi_{\mathrm o} + q_\mathrm{eff} \ell \frac{1 - z}{z} \right)^2 - m_\mathrm{eff}^2 \ell^2 \frac{1 - z^2}{z^2} \right] \phi(z) = 0,
\end{equation}
where the effective charge of the probe field is $q_\mathrm{eff} \equiv (d-2) \frac{\mu \ell}{r_{\mathrm o}} q$. The singularities of eq.~(\ref{eom3}) are located at $z = 0, z = \pm 1$ and $z = \infty$.

To find the solutions, we determine its indices at each singular point. When $z \to 0$, setting $\phi(z) \sim z^{\bar{\alpha}}$, the leading terms in eq.~(\ref{eom3}) are
\begin{equation} \label{eom4}
z^2 \phi''(z) + (q_\mathrm{eff}^2 - m_\mathrm{eff}^2) \ell^2 \phi(z) = 0,
\end{equation}
which gives
\begin{equation}\label{cfm2d}
\bar{\alpha} = \frac12 \pm \frac12 \sqrt{1 + 4 ( m_\mathrm{eff}^2 - q_\mathrm{eff}^2 ) \ell^2} \equiv \frac12 \pm \frac12 \sqrt{1 + 4 \tilde{m}_\mathrm{eff}^2 \ell^2} \equiv \frac12 \pm \nu.
\end{equation}
When $z \to -1$, setting $\phi(z) \sim (1 + z)^{\bar{\beta}}$, then eq.~(\ref{eom3}) reduces to
\begin{equation} \label{eom5}
2 (1 + z) \phi''(z) + 2 \phi'(z) + \frac{(w \xi_{\mathrm o} - 2 q_\mathrm{eff} \ell)^2}{2 (1 + z)} \phi(z) = 0,
\end{equation}
and the index is
\begin{equation}
\bar{\beta} = \pm i \left( \frac{w \xi_{\mathrm o}}{2} - q_\mathrm{eff} \ell \right) = \pm i \left( \frac{w}{4 \pi T_{\mathrm n}} - q_\mathrm{eff} \ell \right).
\end{equation}
Finally, when $z \to 1$, setting $\phi(z) \sim (1 - z)^{\bar{\gamma}}$, then eq.~(\ref{eom3}) reduces to
\begin{equation} \label{eom6}
2 (1 - z) \phi''(z) - 2 \phi'(z) + \frac{(w \xi_{\mathrm o})^2}{2 (1 - z)} \phi(z) = 0,
\end{equation}
from which
\begin{equation}
\bar{\gamma} = \pm i \frac{w \xi_{\mathrm o}}{2} = \pm i \frac{w}{4 \pi T_{\mathrm n}}=\pm i\frac{\omega/\varepsilon}{4\pi /(2\pi \xi_{\mathrm o})}=\pm i\frac{\omega}{2\varepsilon\rho_{\mathrm o}/\ell^2}=\pm i\frac{\omega}{4\pi T}
\end{equation}
is obtained. Further, imposing the ingoing boundary condition at the black brane horizon $z = 1$ requires $\bar{\gamma} = -i \frac{w \xi_{\mathrm o}}{2} = -i \frac{w}{4\pi T_{\mathrm n}}$.

Also note that eq.~(\ref{eom3}) can be rewritten in a more explicit form as
\begin{eqnarray} \label{eom7}
\phi''(z) &+& \left( \frac{1}{z+1} + \frac{1}{z-1} \right) \phi'(z)
\nonumber\\
&+& \left( \frac{\tilde{m}_\mathrm{eff}^2 \ell^2}{z} + \frac{\frac12 (w \xi_{\mathrm o} - 2 q_\mathrm{eff} \ell)^2}{z+1} + \frac{\frac12 w^2 \xi_{\mathrm o}^2}{z-1} \right) \frac{\phi(z)}{z (z+1) (z-1)} = 0,
\end{eqnarray}
which becomes the Fuchs equation with three canonical singularities $a_1$, $a_2$ and $a_3$ as
\begin{eqnarray} \label{fuchs}
&&\phi''(z) + \left( \frac{1-\bar{\alpha}_1-\bar{\alpha}_2}{z-a_1} + \frac{1-\bar{\beta}_1-\bar{\beta}_2}{z-a_2} + \frac{1-\bar{\gamma}_1-\bar{\gamma}_2}{z-a_3} \right) \phi'(z)
\nonumber\\
&&+ \left( \frac{\bar{\alpha}_1 \bar{\alpha}_2 (a_1-a_2) (a_1-a_3)}{z-a_1} + \frac{\bar{\beta}_1 \bar{\beta}_2 (a_2-a_3) (a_2-a_1)}{z-a_2} + \frac{\bar{\gamma}_1 \bar{\gamma}_2 (a_3-a_1) (a_3-a_2)}{z-a_3} \right)\nno\\
&&\times\frac{\phi(z)}{(z-a_1)(z-a_2)(z-a_3)} = 0,
\end{eqnarray}
where $a_1 = 0$, $a_2 = -1$ and $a_3 = 1$ and
\begin{equation}
\bar{\alpha}_1 = \frac12 \pm \nu, \quad \bar{\alpha}_2 = \frac12 \mp \nu, \quad \bar{\beta}_1 = - \bar{\beta}_2 = \pm i \frac{w \xi_{\mathrm o} - 2 q_\mathrm{eff} \ell}2, \quad \bar{\gamma}_1 = - \bar{\gamma}_2 = \pm i \frac{w \xi_{\mathrm o}}2,
\end{equation}
and $\bar{\alpha}_1 + \bar{\alpha}_2 + \bar{\beta}_1 + \bar{\beta}_2 + \bar{\gamma}_1 + \bar{\gamma}_2 = 1$ is satisfied. The Fuchs equation~(\ref{fuchs}) can be transformed to the standard hypergeometric function
\begin{equation}
\zeta (1 - \zeta) \psi''(\zeta) + \left[ \tilde{\gamma} - (1 + \tilde{\alpha} + \tilde{\beta}) \zeta \right] \psi'(\zeta) - \tilde{\alpha} \tilde{\beta} \psi(\zeta) = 0,
\end{equation}
via the conformal coordinate transformation
\begin{equation}
\zeta = \frac{(a_2-a_3)(z-a_1)}{(a_2-a_1)(z-a_3)}, \qquad {\rm and} \qquad \phi(z) = \left(\frac{z-a_1}{z-a_3}\right)^{\bar{\alpha}_1} \left(\frac{z-a_2}{z-a_3}\right)^{\bar{\beta}_1} \psi(\zeta),
\end{equation}
where $\tilde{\alpha} = \bar{\alpha}_1 + \bar{\beta}_1 + \bar{\gamma}_1, \tilde{\beta} = \bar{\alpha}_1 + \bar{\beta}_1 + \bar{\gamma}_2$ and $\tilde{\gamma} = 1 + \bar{\alpha}_1 - \bar{\alpha}_2$. (Note that one can freely choose the indices $i = 1, 2$ for $\bar{\alpha}_i$, $\bar{\beta}_i$ and $\bar{\gamma}_i$.)

For eq.~(\ref{eom7}), we have $\zeta = 2z/(z-1)$ and $\tilde{\alpha} = \frac12 \pm \nu + i w \xi_{\mathrm o} - i q_\mathrm{eff} \ell, \tilde{\beta} = \frac12 \pm \nu - i q_\mathrm{eff} \ell, \tilde{\gamma} = 1 \pm 2\nu$, respectively. Therefore the explicit solutions in the near horizon near extreme region are
\begin{eqnarray} \label{solu}
\phi(z) &=& c_1 \left(\frac{z}{z-1}\right)^{\frac12 + \nu} \left(\frac{z+1}{z-1}\right)^{i \frac{w \xi_{\mathrm o}}{2} - i q_\mathrm{eff} \ell} {_2F_1}\left(\frac12 + \nu + i w \xi_{\mathrm o} - i q_\mathrm{eff} \ell, \frac12 + \nu - i q_\mathrm{eff} \ell; 1 + 2 \nu; \frac{2z}{z-1} \right)
\nonumber\\
&& + c_2 \left(\frac{z}{z-1}\right)^{\frac12 - \nu} \left(\frac{z+1}{z-1}\right)^{i \frac{w \xi_{\mathrm o}}{2} - i q_\mathrm{eff} \ell} {_2F_1}\left(\frac12 - \nu + i w \xi_{\mathrm o} - i q_\mathrm{eff} \ell, \frac12 - \nu - i q_\mathrm{eff} \ell; 1 - 2 \nu; \frac{2z}{z-1} \right).\nno\\
\end{eqnarray}

\subsection{2-point correlators from AdS$_2$}
At the horizon of the AdS$_2$ black brane $z = 1$, eq.~(\ref{solu}) is expanded as
\begin{equation} \label{solu3}
\phi(z) = c_H^{(\mathrm {in})}(1-z)^{-i \frac{w}{4\pi T_{\mathrm n}}} + c_H^{(\mathrm {out})} (1-z)^{i \frac{w}{4\pi T_{\mathrm n}}},
\end{equation}
where
\begin{eqnarray}
c_H^{(\mathrm {in})} &=& c_1 (-)^{-\frac12 - \nu - i \frac{w}{2 \pi T_{\mathrm n}} + i q_\mathrm{eff} \ell} \, 2^{-\frac12 - \nu + i \frac{w}{4\pi T_{\mathrm n}}}
\frac{\Gamma\left(1 + 2\nu\right) \Gamma\left(i \frac{w}{2\pi T_{\mathrm n}}\right)}{\Gamma\left(\frac12 + \nu + i q_\mathrm{eff} \ell\right) \Gamma\left(\frac12 + \nu + i \frac{w}{2\pi T_{\mathrm n}} - i q_\mathrm{eff} \ell\right)}
\nonumber\\
&&+ c_2 (-)^{-\frac12 + \nu - i \frac{w}{2 \pi T_{\mathrm n}} + i q_\mathrm{eff} \ell} \, 2^{-\frac12 + \nu + i \frac{w}{4\pi T_{\mathrm n}}}
\frac{\Gamma\left(1 - 2\nu\right) \Gamma\left(i \frac{w}{2\pi T_{\mathrm n}}\right)}{\Gamma\left(\frac12 - \nu + i q_\mathrm{eff} \ell\right) \Gamma\left(\frac12 - \nu + i \frac{w}{2\pi T_{\mathrm n}} - i q_\mathrm{eff} \ell\right)},
\end{eqnarray}
and
\begin{eqnarray}
c_H^{(\mathrm {out})} &=& c_1 (-)^{-\frac12 - \nu - i \frac{w}{2 \pi T_{\mathrm n}} + i q_\mathrm{eff} \ell} \, 2^{-\frac12 - \nu - i \frac{w}{4 \pi T_{\mathrm n}}}
\frac{\Gamma\left(1 + 2\nu\right) \Gamma\left(-i \frac{w}{2\pi T_{\mathrm n}}\right)}{\Gamma\left(\frac12 + \nu - i q_\mathrm{eff} \ell\right) \Gamma\left(\frac12 + \nu - i \frac{w}{2\pi T_{\mathrm n}} + i q_\mathrm{eff} \ell\right)}
\nonumber\\
&&+ c_2 (-)^{-\frac12 + \nu - i \frac{w}{2 \pi T_{\mathrm n}} + i q_\mathrm{eff} \ell} \, 2^{-\frac12 + \nu - i \frac{w}{4\pi T_{\mathrm n}}}
\frac{\Gamma\left(1 - 2\nu\right) \Gamma\left(-i \frac{w}{2\pi T_{\mathrm n}}\right)}{\Gamma\left(\frac12 - \nu - i q_\mathrm{eff} \ell\right) \Gamma\left(\frac12 - \nu - i \frac{w}{2\pi T_{\mathrm n}} + i q_\mathrm{eff} \ell\right)}.
\end{eqnarray}

On the other hand, at the AdS$_2$ boundary $z \rightarrow 0$, the asymptotic expansion of eq.~(\ref{solu}) is
\begin{equation} \label{solu2}
\phi(z) = c_2 (-)^{\frac12 - \nu + i \frac{w}{4\pi T_{\mathrm n}} - i q_\mathrm{eff} \ell} z^{\frac{1}{2}- \nu} + c_1 (-)^{\frac12 + \nu + i \frac{w}{4\pi T_{\mathrm n}} - i q_\mathrm{eff} \ell} z^{\frac{1}{2}+ \nu}=\mathcal{A}(w, \vec{k})z^{\frac{1}{2}- \nu}+\mathcal{B}(w, \vec{k})z^{\frac{1}{2}+ \nu},
\end{equation}
where $\mathcal{A}$ is the source of the charged scalar field in the bulk AdS$_2$, while $\mathcal{B}$ is the response or the operator $\mathcal{\hat{O}}(w,\vec{k})$ (in the momentum space) of the boundary CFT$_1$ (i.e. the IR CFT) dual to the charged scalar field in the bulk AdS$_2$ background. Note that in order to obtain the propagating modes, $\nu$ should be purely imaginary, which can be set as $\nu \equiv i |\nu|$, i.e. $\phi(z) =  c_B^{(\mathrm {out})} z^{\frac{1}{2} - i|\nu|} + c_B^{(\mathrm {in})} z^{\frac{1}{2} + i|\nu|}$. It was shown in \cite{Chen:2012yn} that the condition of an imaginary $\nu$ is equivalent to the violation of the BF bound in AdS$_2$ spacetime, namely,
\be\label{BF2}
\tilde{m}_\mathrm{eff}^2 < -\frac{1}{4\ell^2},
\ee
which corresponds to a complex conformal weight of the scalar operator in the dual IR CFT.

\subsubsection{Pair production rate and absorption cross section ratio}
The Schwinger pair production rate $|\mathfrak{b}|^2$ and the absorption cross section ratio $\sigma_{\mathrm{abs}}$ can be calculated from the radial flux by imposing different boundary conditions
\be \mathcal{F} = i \left(\frac{r_{\mathrm o}}{L}\right)^{d-1} (1-z^2) (\Phi \partial_z \Phi^* - \Phi^* \partial_z \Phi),
\ee
which gives
\begin{eqnarray}
\mathcal{F}_B^{(\mathrm {in})} = 2 |\nu| \left(\frac{r_{\mathrm o}}{L}\right)^{d-1} |c_B^{(\mathrm {in})}|^2, &\quad& \mathcal{F}_B^{(\mathrm {out})} = -2 |\nu| \left(\frac{r_{\mathrm o}}{L}\right)^{d-1} |c_B^{(\mathrm {out})}|^2,
\nonumber\\
\mathcal{F}_H^{(\mathrm {in})} = \frac{w}{2\pi T_{\mathrm n}} \left(\frac{r_{\mathrm o}}{L}\right)^{d-1} |c_H^{(\mathrm {in})}|^2, &\quad& \mathcal{F}_H^{(\mathrm {out})} = -\frac{w}{2\pi T_{\mathrm n}} \left(\frac{r_{\mathrm o}}{L}\right)^{d-1} |c_H^{(\mathrm {out})}|^2,
\end{eqnarray}
where $\mathcal{F}_B^{(\mathrm {in})}$ and $\mathcal{F}_B^{(\mathrm {out})}$ are the ingoing and outgoing fluxes at the AdS$_2$ boundary, while $\mathcal{F}_H^{(\mathrm {in})}$ and $\mathcal{F}_H^{(\mathrm {out})}$ are the ingoing and outgoing fluxes at the AdS$_2$ black brane horizon, respectively.

The Schwinger pair production rate $|\mathfrak{b}|^2$ can be computed either by choosing the inner boundary condition or the outer boundary conditions, which give the same result~\cite{Chen:2012yn}, e.g., by adopting the outer boundary condition, i.e. $\mathcal{F}_B^{(\mathrm {in})} = 0$, ($c_B^{(\mathrm {in})} = 0 \Rightarrow c_1 = 0$),
\be\label{B}
|\mathfrak{b}|^2&=&\frac{\mathcal{F}_B^{(\mathrm {out})}}{\mathcal{F}_H^{(\mathrm {in})}}=\frac{4 \pi T_{\mathrm n} |\nu|}{w} \left|\frac{c_B^{(\mathrm {out})}}{c_H^{(\mathrm {in})}}\right|^2
\nonumber\\
&=&\frac{ 8 \pi T_{\mathrm n} |\nu|}{w} \left| \frac{\Gamma\left(\frac12 - i|\nu| + i q_\mathrm{eff} \ell\right) \Gamma\left(\frac12 - i|\nu| + i \frac{w}{2\pi T_{\mathrm n}} -i q_\mathrm{eff} \ell\right)}{\Gamma\left(1 - 2i|\nu|\right) \Gamma\left(i \frac{w}{2\pi T_{\mathrm n}}\right)}\right|^2\nno\\
&=&\frac{2\sinh\left(2\pi|\nu| \right)\sinh\left(\frac{w}{2T_{\mathrm n}}\right)}{\cosh\pi\left(|\nu|-q_\mathrm{eff} \ell\right)\cosh\pi\left(|\nu|-\frac{w}{2\pi T_{\mathrm n}}+q_\mathrm{eff} \ell\right)}.
\ee
Similarly, by adopting the outer boundary condition, the absorption cross section ratio is computed as
\begin{eqnarray}\label{A}
\sigma_{\mathrm{abs}} &=& \frac{\mathcal{F}_B^{(\mathrm {out})}}{\mathcal{F}_H^{(\mathrm {out})}} = \frac{4 \pi T_{\mathrm n} |\nu|}{w} \left|\frac{c_B^{(\mathrm {out})}}{c_H^{(\mathrm {out})}}\right|^2
\nonumber\\
&=&\frac{ 8 \pi T_{\mathrm n} |\nu|}{w} \left| \frac{\Gamma\left(\frac12 - i|\nu| - i q_\mathrm{eff} \ell\right) \Gamma\left(\frac12 - i|\nu| - i \frac{w}{2\pi T_{\mathrm n}} +i q_\mathrm{eff} \ell\right)}{\Gamma\left(1 - 2i|\nu|\right) \Gamma\left(-i \frac{w}{2\pi T_{\mathrm n}}\right)}\right|^2\nno\\
&=&\frac{2\sinh\left(2\pi|\nu| \right)\sinh\left(\frac{w}{2T_{\mathrm n}}\right)}{\cosh\pi\left(|\nu|+q_\mathrm{eff} \ell\right)\cosh\pi\left(|\nu|+\frac{w}{2\pi T_{\mathrm n}}-q_\mathrm{eff} \ell\right)}.
\end{eqnarray}
The pair production rate and the absorption cross section ratio are connected by the simple relation
\be\label{abspair2}
|\mathfrak{b}|^2&=&-\sigma_{\mathrm{abs}}(|\nu|\rightarrow -|\nu|),
\ee
which has been observed in~\cite{Chen:2017mnm}.

\subsubsection{Retarded Green's function}
The two-point retarded Green's function of the boundary operators dual to the bulk charged scalar field is computed through
\begin{equation}
G_R^{\mathrm {AdS_2}}(w, \vec{k}) \equiv \langle \mathcal{\hat{O}} \mathcal{\hat{O}} \rangle_R = -2 \mathcal{F}|_{z\rightarrow 0} \sim \frac{\mathcal{B}(w, \vec{k})}{\mathcal{A}(w, \vec{k})} + \textrm{contact terms}
\end{equation}
by taking the inner boundary condition, i.e. $\mathcal{F}_H^{(\mathrm {out})}=0$, which gives
\begin{equation}
\frac{c_2}{c_1} = (-)^{1 - 2\nu} \, 2^{-2\nu} \frac{\Gamma\left(1 + 2\nu\right) \Gamma\left(\frac12 - \nu - i q_\mathrm{eff} \ell\right) \Gamma\left(\frac12 - \nu - i \frac{w}{2\pi T_{\mathrm n}} + i q_\mathrm{eff} \ell\right)}{\Gamma\left(1 - 2\nu\right) \Gamma\left(\frac12 + \nu - i q_\mathrm{eff} \ell\right) \Gamma\left(\frac12 + \nu - i \frac{w}{2\pi T_{\mathrm n}} + i q_\mathrm{eff} \ell\right)}.
\end{equation}
Thus, the two-point retarded Green's function is
\begin{equation}\label{Gr2}
G_R^{\mathrm {AdS_2}}(w, \vec{k}) \sim \frac{\mathcal{B}(\omega, \vec{k})}{\mathcal{A}(\omega, \vec{k})} = (-)^{2\nu} \frac{c_1}{c_2} = (-)^{4\nu-1} \, 2^{2\nu} \frac{\Gamma\left(1 - 2\nu\right) \Gamma\left(\frac12 + \nu - i q_\mathrm{eff} \ell\right) \Gamma\left(\frac12 + \nu - i \frac{w}{2\pi T_{\mathrm n}} + i q_\mathrm{eff} \ell\right)}{\Gamma\left(1 + 2\nu\right) \Gamma\left(\frac12 - \nu - i q_\mathrm{eff} \ell\right) \Gamma\left(\frac12 - \nu - i \frac{w}{2\pi T_{\mathrm n}} + i q_\mathrm{eff} \ell\right)}.
\end{equation}

In addition, the corresponding boundary condition ($\mathcal{F}_B^{(\mathrm {in})}=0$ and $\mathcal{F}_H^{(\mathrm {out})}=0$) is used to obtain the quasinormal modes of the charged scalar field in the AdS$_2$ spacetime, which correspond to the poles of the retarded Green's function of dual operators (with complex conformal weight $h_R=\frac12-\nu$) in the IR CFT, namely,
\be\label{quasi1}
\frac12 - \nu - i \frac{w}{2\pi T_{\mathrm n}} + i q_\mathrm{eff} \ell=-N \Rightarrow w=2\pi T_{\mathrm n}\left(q_\mathrm{eff} \ell-iN-ih_R\right),\quad N=0,1,\cdots.
\ee
Eq.~(\ref{quasi1}) gives the quasinormal modes of the charged scalar field perturbation.

\section{Pair production in the asymptotical AdS$_{5}$} \label{sec:AdS5 pair production}

\subsection{Transforming the radial equation into the Heun-like equation}
To find the solution in the full region, we now focus on $d=4$ and the near extremal cases. Introducing the coordinate transformation $\varrho  = \frac{{{r^2}}}{{M'}}$ (with $M'{\text{ = }}{{\text{G}}_{d + 1}}{L^2}M$), we write the radial equation (\ref{eomtr}) as
\be\label{the primary equation}&&\phi ''(\varrho ) + \left( {\frac{1}{{\varrho  - {\varrho _1}}} + \frac{1}{{\varrho  - {\varrho _2}}} + \frac{1}{{\varrho  - {\varrho_{\mathrm o} }}}} \right)\phi '\left( \varrho  \right) \nno\\
&&+ \left( {\frac{{\varrho {{\left( {\tilde \omega \varrho  - \tilde q\mu {\varrho_{\mathrm o} }} \right)}^2}}}{{{{\left( {\varrho  - {\varrho _1}} \right)}^2}{{\left( {\varrho  - {\varrho _2}} \right)}^2}{{\left( {\varrho  - {\varrho_{\mathrm o} }} \right)}^2}}} - \frac{{{{\tilde m}^{\rm{2}}}\varrho  + {{\tilde k}^2}}}{{\left( {\varrho  - {\varrho _1}} \right)\left( {\varrho  - {\varrho _2}} \right)\left( {\varrho  - {\varrho_{\mathrm o} }} \right)}}} \right)\phi \left( \varrho  \right) = 0,
\ee
where the parameters are $\tilde \omega  = \frac{{{L^2}(\omega  + q\mu )}}{{2\sqrt {M'} }}$, $ \tilde q = \frac{{{L^2}q}}{{2\sqrt {M'} }}$, $\tilde m = \frac{Lm}{2}$, $\tilde k = \frac{{{L^2}\left| {\vec k} \right|}}{{2\sqrt {M'} }}$ and
\be
\varrho _1 &=& - \frac{1}{2}{\varrho_{\mathrm o} } - \frac{1}{2}\sqrt {{\varrho_{\mathrm o} }^{\text{2}} + 8\frac{{{\varrho _*}^{\text{3}}}}{{{\varrho_{\mathrm o} }}}}, \qquad \Bigl(\varrho _*=\frac{{{r_{\text{*}}}^2}}{M'} \Bigr) \nno\\
\varrho _2 &=& - \frac{1}{2}{\varrho_{\mathrm o} }{\text{ + }}\frac{1}{2}\sqrt {{\varrho_{\mathrm o} }^{\text{2}} + 8\frac{{{\varrho _*}^{\text{3}}}}{{{\varrho_{\mathrm o} }}}}.
\ee
Further, defining another coordinate
\be
y\equiv \frac{{\varrho  - {\varrho_{\mathrm o} }}}{{{\varrho_{\mathrm o} }}},\quad a\equiv  \frac{\varrho _2-\varrho _{\mathrm{o}}}{\varrho _{\mathrm{o}}},\quad {\rm and}\quad b\equiv  \frac{\varrho _1-\varrho _{\mathrm{ o }}}{\varrho _{\mathrm{ o }}},
\ee
the metric of the RN-AdS$_5$ black hole becomes
\be\label{RNAdSy}
ds^2&=&\frac{L^2 dy^2}{4(1+y)^2 f(y)}+\frac{r_{\mathrm{o}}^2}{L^2}(1+y)\left(-f(y)dt^2+dx_i^2 \right),\nno\\
A&=&\frac{\mu y}{1+y}dt,
\ee
where
\be
f(y)=1-\frac{M'}{r_{\mathrm{o}}^4}(1+y)^{-2}+\frac{Q'^2}{r_{\mathrm{o}}^6}(1+y)^{-3}, \qquad
{{Q'}^2} = {{\text{G}}_{d + 1}}{L^2}{Q^2}.
\ee
And eq.~(\ref{the primary equation}) transforms into
\be\label{equ2}\phi ''(y) + \left( {\frac{1}{y} + \frac{1}{{y - a}} + \frac{1}{{y - b}}} \right)\phi '\left( y \right) + \left( {\frac{{{{\left( {\tilde \omega (y + 1) - \tilde q\mu } \right)}^2}(y + 1)}}{{{y^2}{{\left( {y - a} \right)}^2}{{\left( {y - b} \right)}^2}}} - \frac{{{{\tilde m}^2}(y + 1){\varrho_{\mathrm o} } + {{\tilde k}^2}}}{{y\left( {y - a} \right)\left( {y - b} \right)}}} \right)\frac{{\phi \left( y \right)}}{{{\varrho_{\mathrm o} }}} = 0.\nno\\
\ee
We employ the Riemann's P-function~\cite{Hall:1995} for the solution
\be
\phi \left( y \right) = P\left( \begin{array}{l}
\begin{array}{*{20}{c}}
0&{{\kern 1pt} {\kern 1pt} {\kern 1pt} {\kern 1pt} {\kern 1pt} a}&{{\kern 1pt} {\kern 1pt} {\kern 1pt} {\kern 1pt} {\kern 1pt} {\kern 1pt} b}&{{\kern 1pt} {\kern 1pt} {\kern 1pt} {\kern 1pt} {\kern 1pt} {\kern 1pt} {\kern 1pt} \infty }
\end{array}\\
\begin{array}{*{20}{c}}
{{\alpha _1}}&{{\gamma _1}}&{{\beta _1}}&{{\kern 1pt} {\kern 1pt} {\delta _1}{\kern 1pt} {\kern 1pt} {\kern 1pt} {\kern 1pt} {\kern 1pt} {\kern 1pt} {\kern 1pt} {\kern 1pt} {\kern 1pt} {\kern 1pt} y{\kern 1pt} {\kern 1pt} }
\end{array}\\
\begin{array}{*{20}{c}}
{{\alpha _2}}&{{\gamma _2}}&{{\beta _2}}&{{\delta _2}}
\end{array}
\end{array} \right)
\ee
where $\alpha _{1,2}$, $\beta _{1,2}$, $\gamma _{1,2}$ and $\delta _{1,2}$ are exponents of corresponding singularities $0$, $a$, $b$ and $\infty$, respectively.
\be
&&\alpha _{1,2} =  \pm i\frac{(\tilde \omega  - \tilde q\mu) }{ab\sqrt {\varrho_{\mathrm o}}}=\pm\frac{i\omega}{4\pi T},\nno\\
&&\beta _{1,2} =  \pm i\frac{{(\tilde \omega (1 + b) - \tilde q\mu )\sqrt {1 + b} }}{{(a - b)b\sqrt {{\varrho_{\mathrm o} }} }},\quad \gamma _{1,2} =  \pm i\frac{{(\tilde \omega (1 + a) - \tilde q\mu )\sqrt {1 + a} }}{{(b - a)a\sqrt {{\varrho_{\mathrm o} }} }},
\ee
in which the index ``1'' corresponds to the $``+''$ sign and the index ``2'' corresponds to the $``-''$ sign, respectively.

By using the transformation identity of the P-function, one can write the solution as
\be \label{equ3}
\phi \left( y \right) = {\left( {\frac{y}{{y - b}}} \right)^{{\alpha _1}}} R(y)
\ee
where
\be
R(y)\equiv P\left( \begin{array}{l}
\begin{array}{*{20}{c}}
{{\kern 1pt} {\kern 1pt} {\kern 1pt} {\kern 1pt} {\kern 1pt} {\kern 1pt} {\kern 1pt} {\kern 1pt} {\kern 1pt} {\kern 1pt} {\kern 1pt} {\kern 1pt} {\kern 1pt} 0}&{{\kern 1pt} {\kern 1pt} {\kern 1pt} {\kern 1pt} {\kern 1pt} {\kern 1pt} {\kern 1pt} {\kern 1pt} {\kern 1pt} {\kern 1pt} {\kern 1pt} {\kern 1pt} {\kern 1pt} {\kern 1pt} {\kern 1pt} {\kern 1pt} {\kern 1pt} {\kern 1pt} {\kern 1pt} a}&{{\kern 1pt} {\kern 1pt} {\kern 1pt} {\kern 1pt} {\kern 1pt} {\kern 1pt} {\kern 1pt} {\kern 1pt} {\kern 1pt} {\kern 1pt} {\kern 1pt} {\kern 1pt} {\kern 1pt} {\kern 1pt} {\kern 1pt} {\kern 1pt} {\kern 1pt} {\kern 1pt} b}&{{\kern 1pt} {\kern 1pt} {\kern 1pt} {\kern 1pt} {\kern 1pt} {\kern 1pt} {\kern 1pt} {\kern 1pt} {\kern 1pt} {\kern 1pt} {\kern 1pt} {\kern 1pt} {\kern 1pt} {\kern 1pt} {\kern 1pt} \infty }
\end{array}\\
{\kern 1pt} {\kern 1pt} {\kern 1pt} {\kern 1pt} {\kern 1pt} {\kern 1pt} {\kern 1pt} {\kern 1pt} {\kern 1pt} {\kern 1pt} {\kern 1pt} {\kern 1pt} \begin{array}{*{20}{c}}
{{\kern 1pt} 0}&{{\kern 1pt} {\kern 1pt} {\kern 1pt} {\kern 1pt} {\kern 1pt} {\kern 1pt} {\kern 1pt} {\kern 1pt} {\kern 1pt} {\kern 1pt} {\kern 1pt} {\kern 1pt} {\kern 1pt} {\kern 1pt} {\kern 1pt} {\kern 1pt} {\kern 1pt} {\gamma _1}}&{{\kern 1pt} {\kern 1pt} {\beta _1} + {\alpha _1}}&{{\kern 1pt} {\delta _1}{\kern 1pt} {\kern 1pt} {\kern 1pt} {\kern 1pt} {\kern 1pt} {\kern 1pt} {\kern 1pt} {\kern 1pt} {\kern 1pt} {\kern 1pt} y}
\end{array}\\
\begin{array}{*{20}{c}}
{{\alpha _2} - {\alpha _1}}&{{\kern 1pt} {\gamma _2}}&{{\beta _2} + {\alpha _1}}&{{\kern 1pt} {\kern 1pt} {\delta _2}}
\end{array}
\end{array} \right).
\ee
Substituting eq.~(\ref{equ3}) into eq.~(\ref{equ2}), we obtain
\be\label{equ4}R''(y) + \left( {\frac{1}{y} + \frac{1}{{y - a}} + \frac{1}{{y - b}} - \frac{{2b{\alpha _1}}}{{y(y - b)}}} \right)R'\left( y \right) + {V_1}R(y) = 0\ee
where
\be
V_1 &\equiv & \frac{{{a^2}{{\left( {\tilde \omega (y + 1) - \tilde q\mu } \right)}^2} + {a^2}\left( {\tilde \omega y + 2(\tilde \omega  - \tilde q\mu )} \right)\tilde \omega  - (y - 2a){{(\tilde \omega  - \tilde q\mu )}^2}}}{{{a^2}{\varrho_{\mathrm o} }y{{\left( {y - a} \right)}^2}{{\left( {y - b} \right)}^2}}}\nno\\
&&- \frac{{{{\tilde m}^2}(y + 1){\varrho_{\mathrm o} } + {{\tilde k}^2}}}{{{\varrho_{\mathrm o} }y\left( {y - a} \right)\left( {y - b} \right)}} - \frac{{b{\alpha _1}}}{{y\left( {y - a} \right)(y - b)}}.
\ee
Further, using another identity of the P-function, the solution is transformed into
\be\label{equ5}
R(y) = {\left( {\frac{{y - a}}{{y - b}}} \right)^{{\gamma _1}}} T(y)
\ee
where
\be
T(y)\equiv P\left( \begin{array}{l}
\begin{array}{*{20}{c}}
{{\kern 1pt} {\kern 1pt} {\kern 1pt} {\kern 1pt} {\kern 1pt} {\kern 1pt} {\kern 1pt} {\kern 1pt} {\kern 1pt} {\kern 1pt} {\kern 1pt} {\kern 1pt} {\kern 1pt} 0}&{{\kern 1pt} {\kern 1pt} {\kern 1pt} {\kern 1pt} {\kern 1pt} {\kern 1pt} {\kern 1pt} {\kern 1pt} {\kern 1pt} {\kern 1pt} {\kern 1pt} {\kern 1pt} {\kern 1pt} {\kern 1pt} {\kern 1pt} {\kern 1pt} {\kern 1pt} {\kern 1pt} {\kern 1pt} {\kern 1pt} {\kern 1pt} {\kern 1pt} {\kern 1pt} {\kern 1pt} {\kern 1pt} {\kern 1pt} {\kern 1pt} {\kern 1pt} {\kern 1pt} {\kern 1pt} {\kern 1pt} {\kern 1pt} a}&{{\kern 1pt} {\kern 1pt} {\kern 1pt} {\kern 1pt} {\kern 1pt} {\kern 1pt} {\kern 1pt} {\kern 1pt} {\kern 1pt} {\kern 1pt} {\kern 1pt} {\kern 1pt} {\kern 1pt} {\kern 1pt} {\kern 1pt} {\kern 1pt} {\kern 1pt} {\kern 1pt} {\kern 1pt} {\kern 1pt} {\kern 1pt} {\kern 1pt} {\kern 1pt} {\kern 1pt} {\kern 1pt} {\kern 1pt} {\kern 1pt} {\kern 1pt} {\kern 1pt} {\kern 1pt} {\kern 1pt} {\kern 1pt} {\kern 1pt} {\kern 1pt} {\kern 1pt} {\kern 1pt} b}&{{\kern 1pt} {\kern 1pt} {\kern 1pt} {\kern 1pt} {\kern 1pt} {\kern 1pt} {\kern 1pt} {\kern 1pt} {\kern 1pt} {\kern 1pt} {\kern 1pt} {\kern 1pt} {\kern 1pt} {\kern 1pt} {\kern 1pt} {\kern 1pt} {\kern 1pt} {\kern 1pt} {\kern 1pt} {\kern 1pt} {\kern 1pt} {\kern 1pt} {\kern 1pt} {\kern 1pt} {\kern 1pt} {\kern 1pt} \infty }
\end{array}\\
{\kern 1pt} {\kern 1pt} {\kern 1pt} {\kern 1pt} {\kern 1pt} {\kern 1pt} {\kern 1pt} {\kern 1pt} {\kern 1pt} {\kern 1pt} {\kern 1pt} {\kern 1pt} \begin{array}{*{20}{c}}
{{\kern 1pt} 0}&{{\kern 1pt} {\kern 1pt} {\kern 1pt} {\kern 1pt} {\kern 1pt} {\kern 1pt} {\kern 1pt} {\kern 1pt} {\kern 1pt} {\kern 1pt} {\kern 1pt} {\kern 1pt} {\kern 1pt} {\kern 1pt} {\kern 1pt} {\kern 1pt} {\kern 1pt} {\kern 1pt} {\kern 1pt} {\kern 1pt} {\kern 1pt} {\kern 1pt} {\kern 1pt} {\kern 1pt} {\kern 1pt} {\kern 1pt} {\kern 1pt} {\kern 1pt} {\kern 1pt} {\kern 1pt} {\kern 1pt} 0}&{{\kern 1pt} {\kern 1pt} {\kern 1pt} {\kern 1pt} {\kern 1pt} {\kern 1pt} {\kern 1pt} {\kern 1pt} {\kern 1pt} {\kern 1pt} {\kern 1pt} {\kern 1pt} {\kern 1pt} {\beta _1} + {\alpha _1} + {\gamma _1}}&{{\kern 1pt} {\delta _1}{\kern 1pt} {\kern 1pt} {\kern 1pt} {\kern 1pt} {\kern 1pt} {\kern 1pt} {\kern 1pt} {\kern 1pt} {\kern 1pt} {\kern 1pt} y}
\end{array}\\
\begin{array}{*{20}{c}}
{{\alpha _2} - {\alpha _1}}&{{\kern 1pt} {\gamma _2} - {\gamma _1}}&{{\beta _2} + {\alpha _1} + {\gamma _1}}&{{\kern 1pt} {\kern 1pt} {\delta _2}}
\end{array}
\end{array} \right).
\ee
Similarly, substituting eq.~(\ref{equ5}) into eq.~(\ref{equ4}), we find
\be\label{equ6}T''(y) + \left( {\frac{1}{y} + \frac{1}{{y - a}} + \frac{1}{{y - b}} - \frac{{2b{\alpha _1}}}{{y(y - b)}} + \frac{{2(a - b){\gamma _1}}}{{(y - a)(y - b)}}} \right)T'\left( y \right) + {V_2}T(y) = 0\ee
where
\be
{V_2} \equiv  - \frac{{(2 + 3a - {a^2}y){{\tilde \omega }^{\rm{2}}} - {\rm{4}}\left( {a + 1} \right)\tilde \omega \tilde q\mu {\rm{ + }}\left( {a + {\rm{2}}} \right){{\tilde q}^{\rm{2}}}{\mu ^{\rm{2}}}}}{{{\varrho _{\rm{o}}}{a^2}y\left( {y - a} \right){{\left( {y - b} \right)}^2}}} - {M_1}
\ee
and
\be
M_1= \frac{{\left( {b - a} \right){\gamma _1}}}{{y\left( {y - a} \right)(y - b)}} + \frac{{2b\left( {a - b} \right){\alpha _1}{\gamma _1}}}{{y\left( {y - a} \right){{(y - b)}^2}}} + \frac{{{{\tilde m}^2}(y + 1){\varrho_{\mathrm o} } + {{\tilde k}^2}}}{{{\varrho_{\mathrm o} }y\left( {y - a} \right)\left( {y - b} \right)}} + \frac{{b{\alpha _1}}}{{y\left( {y - a} \right)(y - b)}}.
\ee
So far we have succeeded in transforming one of the exponents of singularities at ``$y=0$'' and ``$y=a$'' into zero respectively by separating the asymptotic factors. Strictly speaking, eq.~(\ref{equ6}) is not the standard Heun's function unless one of the exponents of singularity at ``$y=b$'' also turns into zero (so we will call it the Heun-like equation). Nevertheless, we will show that eq.~(\ref{equ6}) can be analytically solved without the need of further transforming one of the exponents of singularity at ``$y=b$'' into zero.
\subsection{Solutions in near and far regions }
We divide the regions into a near region
\be
y = \frac{{\varrho  - {\varrho_{\mathrm o} }}}{{{\varrho_{\mathrm o} }}} \ll 1,
\ee
and a far region
\be
y = \frac{{\varrho  - {\varrho_{\mathrm o} }}}{{{\varrho_{\mathrm o} }}}\gg -a,
\ee
and an overlapping region when
\be
-a  \ll 1.
\ee
A physical reasoning of $-a \ll 1$ relies on an observation that the temperature of black hole
\be
T = \frac{{{r_{\mathrm o} }}}{{\pi {L^2}}}\left( {1 - \frac{{{\varrho _*}^3}}{{{\varrho_{\mathrm o} }^3}}} \right),
\ee
thus gives $-a \to 0$ as when  $T \to 0$:
\be
- a = \frac{3}{2} - \frac{1}{2}\sqrt {9 - \frac{8\pi L^2 T}{r_{\mathrm o} }}.
\ee

Now we find the approximate solutions in different regions. First, by using the near region condition ($y \ll 1$), eq.~(\ref{equ2}) reduces to
\be\label{equ7}
\phi ''(y) + \left( {\frac{1}{y} + \frac{1}{{y - a}}} \right)\phi '\left( y \right) + \left( { - \frac{{{{(a{\alpha _1} - i{q_{{\rm{eff}}}}\ell y)}^2}}}{{{y^{\rm{2}}}{{\left( {y - a} \right)}^{\rm{2}}}}} + \frac{{{{\tilde m}^2}{\varrho _{\rm{o}}} + {{\tilde k}^2}}}{{{\varrho _{\rm{o}}}by(y - a)}}} \right)\phi (y) = 0
\ee
Obviously eq.~(\ref{equ7}) can be solved by the hypergeometric function as
\be\label{equ8}
\phi \left( y \right) = {\left( {y - a} \right)^{{\alpha _1} - i{q_{{\rm{eff}}}}\ell}}\bigg( {{c_3}{y^{{\alpha _1}}}_2{F_1}\left( {\alpha ,\beta ;\gamma ;\frac{y}{a}} \right) + {c_4}{y^ - }^{{\alpha _1}}{}_2{F_1}\left( {1 - \gamma  + \alpha ,1 - \gamma  + \beta ;2 - \gamma ;\frac{y}{a}} \right)} \bigg),
\ee
where $\alpha  = \frac{1}{2} + \nu  + 2{\alpha _1} - i{q_{{\rm{eff}}}}\ell $, $\beta  = \frac{1}{2} - \nu  + 2{\alpha _1} - i{q_{{\rm{eff}}}}\ell $, $\gamma  = 1 + 2{\alpha _1}$ and $\ell=L/\sqrt{12}$ is the  radius of the effective AdS$_2$ geometry in the near horizon region of the RN-AdS$_5$ black hole.
Second, in the far region, by using the condition ($y \gg -a$), eq.~(\ref{equ6}) can turn into
\be\label{equ9}T''(y) + \left( {\frac{2}{y} + \frac{1}{{y - b}} - \frac{{4b{\alpha _1}}}{{y(y - b)}}{\rm{ + }}\frac{{2ib{q_{{\rm{eff}}}}\ell }}{{y(y - b)}}} \right)T'\left( y \right) + {V_3}T(y) = 0\ee
where
\be
{V_3} \equiv \frac{{{{\tilde \omega }^2}}}{{{\varrho _{\rm{o}}}y{{(y - b)}^2}}} + \frac{{4{b^2}{\alpha _1}\left( {{\alpha _1} - i{q_{{\rm{eff}}}}\ell } \right)}}{{{y^2}{{(y - b)}^2}}} - \frac{{b\left( {2{\alpha _1} - i{q_{{\rm{eff}}}}\ell } \right)}}{{{y^2}(y - b)}} - \frac{{{{\tilde m}^2}(y + 1){\varrho _{\rm{o}}} + {{\tilde k}^2}}}{{{\varrho _{\rm{o}}}{y^2}(y - b)}},
\ee
(where the relation ${\gamma _1} \approx {\alpha _1} + \frac{{i\tilde q\mu }}{{b\sqrt {{\varrho _{\rm{o}}}} }} = {\alpha _1} - i{q_{{\rm{eff}}}}\ell $ when $\left| a \right| \ll 1$ is used). Similarly, eq.~(\ref{equ9}) has a solution in terms of hypergeometric function as
\be\label{equ10}\phi \left( y \right) = {\left( {\frac{y}{b} - 1} \right)^\lambda}\bigg( {{c_5}{y^{\nu  - \frac{1}{2}}}_2{F_1}\left( {\alpha ',\beta ';\gamma ';\frac{y}{b}} \right) + {c_6}{y^{ - \nu  - \frac{1}{2}}}_2{F_1}\left( {1 - \gamma ' + \alpha ',1 - \gamma ' + \beta ';2 - \gamma ';\frac{y}{b}} \right)} \bigg)
\ee
in which $\alpha ' = \frac{1}{2} + \nu  + \Delta  + \lambda$, $\beta ' = \frac{1}{2} + \nu  - \Delta  + \lambda$, $\gamma ' = 1 + 2\nu $, $\Delta  = \sqrt {1 + {{\tilde m}^2}}$ and $\lambda = \sqrt {{{\left( {i{q_{{\rm{eff}}}}\ell } \right)}^2} - \frac{{{{\tilde \omega }^2}}}{{{\varrho _{\rm{o}}}b}}} $.

\subsection{Near-far matching }
In the overlapping region, one has the inequalities
\be
 - a \ll y \ll 1 < -b
 \ee
($1 <  - b$ since $ - b = \frac{3}{2}{\text{ + }}\frac{1}{2}\sqrt {9 - \frac{{8\pi {L^2}T}}{{{r_{\mathrm o} }}}}  \to {\text{3}}$, as $T \to {\text{0}}$), which means $\left| {\frac{a}{y}} \right| \to 0$ and $\left| {\frac{y}{b}} \right| \to 0$, and transforms eqs.~(\ref{equ8}) and (\ref{equ10}) into the following forms: the near region solution
\be
\phi (y) &=& \left( {{{\left( { - 1} \right)}^{ - \alpha }}{c_3}\frac{{\Gamma \left( \gamma  \right)\Gamma \left( {\beta  - \alpha } \right)}}{{\Gamma \left( \beta  \right)\Gamma \left( {\gamma  - \alpha } \right)}} + {{\left( { - 1} \right)}^{ - 1 + \gamma  - \alpha }}{c_4}\frac{{\Gamma \left( {2 - \gamma } \right)\Gamma \left( {\beta  - \alpha } \right)}}{{\Gamma \left( {1 - \gamma  + \beta } \right)\Gamma \left( {1 - \alpha } \right)}}} \right){y^{ - \frac{1}{2} - \nu }}\nno\\
&&+ \left( {{{\left( { - 1} \right)}^{ - \beta }}{c_3}\frac{{\Gamma \left( \gamma  \right)\Gamma \left( {\alpha  - \beta } \right)}}{{\Gamma \left( \alpha  \right)\Gamma \left( {\gamma  - \beta } \right)}} + {{\left( { - 1} \right)}^{ - 1 + \gamma  - \beta }}{c_4}\frac{{\Gamma \left( {2 - \gamma } \right)\Gamma \left( {\alpha  - \beta } \right)}}{{\Gamma \left( {1 - \gamma  + \alpha } \right)\Gamma \left( {1 - \beta } \right)}}} \right){y^{ - \frac{1}{2} + \nu }}
\ee
and the far region solution
\be
\phi (y) \to {\left( { - 1} \right)^\lambda}{c_5}{y^{ - \frac{1}{2} + \nu }} + {\left( { - 1} \right)^\lambda}{c_6}{y^{ - \frac{1}{2} - \nu }}.
\ee
Comparing the above two identities, one finds the connection relations
\be\label{equ11}{c_5} = {( - 1)^{ - \lambda}}\left( {{{\left( { - 1} \right)}^{ - \beta }}{c_3}\frac{{\Gamma \left( \gamma  \right)\Gamma \left( {\alpha  - \beta } \right)}}{{\Gamma \left( \alpha  \right)\Gamma \left( {\gamma  - \beta } \right)}} + {{\left( { - 1} \right)}^{ - 1 + \gamma  - \beta }}{c_4}\frac{{\Gamma \left( {2 - \gamma } \right)\Gamma \left( {\alpha  - \beta } \right)}}{{\Gamma \left( {1 - \gamma  + \alpha } \right)\Gamma \left( {1 - \beta } \right)}}} \right),\ee
\be\label{equ12}{c_6} = {( - 1)^{ - \lambda}}\left( {{{\left( { - 1} \right)}^{ - \alpha }}{c_3}\frac{{\Gamma \left( \gamma  \right)\Gamma \left( {\beta  - \alpha } \right)}}{{\Gamma \left( \beta  \right)\Gamma \left( {\gamma  - \alpha } \right)}} + {{\left( { - 1} \right)}^{ - 1 + \gamma  - \alpha }}{c_4}\frac{{\Gamma \left( {2 - \gamma } \right)\Gamma \left( {\beta  - \alpha } \right)}}{{\Gamma \left( {1 - \gamma  + \beta } \right)\Gamma \left( {1 - \alpha } \right)}}} \right).\ee

\subsection{2-point correlators from AdS$_5$}
\subsubsection{Pair production and absorption cross section}
Now we denote the radial flux of the charged scalar field in metric (\ref{RNAdSy}) as $\mathcal{D}$:
\be\label{fluxads5}
\mathcal{D}=\frac{2ir_{\mathrm{o}}^4(1+y)^3 f(y)}{L^5}\bigg(\phi(y)\partial_y \phi^*(y)- \phi^*(y)\partial_y \phi(y) \bigg).
\ee
In the near horizon limit, i.e. $y \to 0$, eq.~(\ref{equ8}) reduces to
\be
\phi (y) = {c_3}{y^{{\alpha _1}}}{\left( {y - a} \right)^{{\alpha _1} - i{q_{{\rm{eff}}}}\ell }} + {c_4}{y^{ - {\alpha _1}}}{\left( {y - a} \right)^{{\alpha _1} - i{q_{{\rm{eff}}}}\ell }},
\ee
where the first part is the outgoing mode and the second part is the ingoing mode. Further, the asymptotic form of $\phi(y)$ at the boundary ($y\to \infty$) of the AdS$_5$ spacetime gives the form
\be\label{phibdy}
\phi(y)= A(\tilde{\omega}, \tilde{k})y^{- 1 + \Delta}+B(\tilde{\omega}, \tilde{k})y^{- 1 - \Delta},
\ee
where $A(\tilde{\omega}, \tilde{k})$ is the source of the charged scalar field in the bulk RN-AdS$_5$ black hole, while $B(\tilde{\omega}, \tilde{k})$ is the response (the operator) of the boundary CFT$_4$ (i.e. the UV CFT) dual to the charged scalar field in the bulk. As in the case of the AdS$_2$ spacetime, the condition for the propagating modes requires an imaginary $\Delta$, i.e. $\Delta  = i\left| \Delta  \right|$, which means
\be\label{BF5}
m^2\leq -\frac{4}{L^2},
\ee
namely, the violation of the BF bound in AdS$_5$ spacetime.

Therefore, the corresponding outgoing and ingoing fluxes at the horizon and the boundary of the near extremal RN-AdS$_5$ black brane are
\begin{eqnarray}
\mathcal{D}_H^{(\mathrm{out})} &=&\frac{2r_{\mathrm{o}}^3\omega}{L^3}|c_3|^2 =\frac{{4\pi \omega T{r_{\mathrm{o}}}^2}}{{abL}}|{c_3}{|^2},\quad {\kern 1pt} \mathcal{D}_H^{(\mathrm{in})} =-\frac{2r_{\mathrm{o}}^3\omega}{L^3}|c_4|^2  = - \frac{{4\pi \omega T{r_{\rm{o}}}^2}}{{abL}}|{c_4}{|^2}
\nonumber\\
\mathcal{D}_B^{({\mathrm{out}})} &=& \frac{{4|\Delta |{r_{\rm{o}}}^4}}{{{L^5}}}\left| A( {\tilde \omega ,\tilde k})\right|^2,\quad \mathcal{D}_B^{({\mathrm{in}})} =  - \frac{{4|\Delta |{r_{\rm{o}}}^4}}{{{L^5}}}\left|B( {\tilde \omega ,\tilde k})\right|^2.
\end{eqnarray}
The absorption cross section ratio $\sigma _{\mathrm{abs}}^{\mathrm{AdS_5}}$ and the Schwinger pair production rate $\left|\mathfrak{b}^{\mathrm{AdS_5}}\right|^2$ can be calculated by choosing the inner boundary condition $\mathcal{D}_H^{(\mathrm{out})}=0$, ( $c_3=0$), which are given by
\be\label{A5}
\sigma _{{\rm{abs}}}^{{\rm{Ad}}{{\rm{S}}_{\rm{5}}}} = \left| {\frac{{{\cal D}_H^{({\rm{in}})}}}{{{\cal D}_B^{({\rm{in}})}}}} \right| = \frac{{{\rm{2}}T{L^{\rm{2}}}\left| \nu  \right|\sinh \left( {2\pi \left| \Delta  \right|} \right)}}{{{r_{\rm{o}}}{{\left| {H\left( {\nu ;\Delta ;\lambda } \right){{\left( {G_R^{{\rm{Ad}}{{\rm{S}}_{\rm{2}}}}} \right)}^{ - 1}} + H\left( { - \nu ;\Delta ;\lambda } \right)} \right|}^2}}}{\sigma _{{\rm{abs}}}},
\ee
and
\be\label{pair5}
{\left| {{\mathfrak{b}^{{\text{Ad}}{{\text{S}}_{\text{5}}}}}} \right|^2} = \left| {\frac{{\mathcal{D}_H^{({\text{in}})}}}{{\mathcal{D}_B^{({\text{out}})}}}} \right| = \frac{{{\text{2}}T{L^{\text{2}}}\left| \nu  \right|\sinh \left( {2\pi \left| \Delta  \right|} \right)}}{{{r_{\rm{o}}}{{\left| {H\left( { - \nu ; - \Delta ;\lambda } \right)G_R^{{\text{Ad}}{{\text{S}}_{\text{2}}}} + H\left( {\nu ; - \Delta ;\lambda } \right)} \right|}^2}}}{\left| \mathfrak{b} \right|^2},
\ee
where $H\left( {x;y;z} \right)$ denotes a function
\be
H\left( {x;y;z} \right) \equiv {\left( { - 1} \right)^{2x}}{2^x}\frac{{\Gamma \left( {1 + {\rm{2}}x} \right)}}{{\Gamma \left( {\frac{1}{2} + x - y + z} \right)\Gamma \left( {\frac{1}{2} + x - y - z} \right)}}.
\ee
and
\be\label{Gr22}
G_R^{\rm{AdS_2}}={\left( { - 1} \right)^{4\nu  - 1}}{2^{2\nu }}\frac{{\Gamma \left( {1 - 2\nu } \right)}}{{\Gamma \left( {1 + 2\nu } \right)}}\frac{{\Gamma \left( {\frac{1}{2} + \nu  - i{q_{{\rm{eff}}}}\ell } \right)}}{{\Gamma \left( {\frac{1}{2} - \nu  - i{q_{{\rm{eff}}}}\ell } \right)}}\frac{{\Gamma \left( {\frac{1}{2} + \nu  - i\frac{\omega }{{2\pi T}} + i{q_{{\rm{eff}}}}\ell } \right)}}{{\Gamma \left( {\frac{1}{2} - \nu  - i\frac{\omega }{{2\pi T}}{\rm{ + }}i{q_{{\rm{eff}}}}\ell } \right)}},
\ee
which is exactly the retarded Green's function eq.~(\ref{Gr2}) of the IR CFT in the near horizon, near extremal region.
Furthermore, ${\sigma _{\mathrm{abs}}}$ and ${\left| \mathfrak{b} \right|^2}$ are exactly the absorption cross section ratio and the Schwinger pair production rate of the corresponding IR CFT obtained from eqs.~(\ref{A}) and (\ref{B}). We find a relationship
\be\label{abspair5}
\left|\mathfrak{b}^{\mathrm{AdS_5}}\right|^2=-\sigma _{\mathrm{abs}}^{\mathrm{AdS_5}}\left(\left| \nu  \right| \to  - \left| \nu  \right|, \left| \Delta  \right| \to  - \left| \Delta  \right|\right),
\ee
which is similar to eq.~(\ref{abspair2}) except for a combined change of signs in both $|\nu|$ and $|\Delta|$. It can be shown from eqs.~(\ref{A5}) and (\ref{pair5}) that both the absorption cross section ratio $\sigma _{\mathrm{abs}}^{\mathrm{AdS_5}}$ and the pair production ratio $\left|\mathfrak{b}^{\mathrm{AdS_5}}\right|^2$ are proportional to their corresponding counterparts in the AdS$_2$ spacetime in the near horizon, near extremal limit.

\subsubsection{Retarded Green's function}
To calculate the retarded Green's function, the ingoing boundary condition is required, namely, $c_3= 0$. Then from eqs.~(\ref{equ11}) and (\ref{equ12}) the connection relations are
\be\label{equ14}
{c_5} = {\left( { - 1} \right)^{ - 1 + \gamma  - \beta  - \lambda}}{c_4}\frac{{\Gamma \left( {2 - \gamma } \right)\Gamma \left( {\alpha  - \beta } \right)}}{{\Gamma \left( {1 - \gamma  + \alpha } \right)\Gamma \left( {1 - \beta } \right)}}\ee
\be\label{equ15}{c_6} = {\left( { - 1} \right)^{ - 1 + \gamma  - \alpha  - \lambda}}{c_4}\frac{{\Gamma \left( {2 - \gamma } \right)\Gamma \left( {\beta  - \alpha } \right)}}{{\Gamma \left( {1 - \gamma  + \beta } \right)\Gamma \left( {1 - \alpha } \right)}}.
\ee
Substituting eqs.~(\ref{equ14})(\ref{equ15}) into eq.~(\ref{equ10}) and taking $y \to \infty $, namely, the boundary of the AdS$_5$ spacetime, one obtains
\be
A(\tilde \omega ,\tilde k) &=& \left( - 1 \right)^{i q_{\rm{eff}}\ell  - 2\lambda - 1 + \Delta }c_4\bigg( \frac{{\Gamma \left( {2 - \gamma } \right)\Gamma \left( {\alpha  - \beta } \right)}}{{\Gamma \left( {1 - \gamma  + \alpha } \right)\Gamma \left( {1 - \beta } \right)}}\frac{{\Gamma \left( {\gamma '} \right)\Gamma \left( {\alpha ' - \beta '} \right)}}{{\Gamma \left( {\alpha '} \right)\Gamma \left( {\gamma ' - \beta '} \right)}}\nno\\
&&+ \frac{{\Gamma \left( {2 - \gamma } \right)\Gamma \left( {\beta  - \alpha } \right)}}{{\Gamma \left( {1 - \gamma  + \beta } \right)\Gamma \left( {1 - \alpha } \right)}}\frac{{\Gamma \left( {2 - \gamma '} \right)\Gamma \left( {\alpha ' - \beta '} \right)}}{{\Gamma \left( {1 - \gamma ' + \alpha '} \right)\Gamma \left( {1 - \beta '} \right)}} \bigg), \nno\\
B(\tilde{\omega}, \tilde{k})&=& \left( - 1 \right)^{i q_{\rm{eff}}\ell  - 2\lambda - 1 - \Delta }c_4\bigg( \frac{{\Gamma \left( {2 - \gamma } \right)\Gamma \left( {\alpha  - \beta } \right)}}{{\Gamma \left( {1 - \gamma  + \alpha } \right)\Gamma \left( {1 - \beta } \right)}}\frac{{\Gamma \left( {\gamma '} \right)\Gamma \left( {\beta ' - \alpha '} \right)}}{{\Gamma \left( {\beta '} \right)\Gamma \left( {\gamma ' - \alpha '} \right)}}\nno\\
&&+ \frac{{\Gamma \left( {2 - \gamma } \right)\Gamma \left( {\beta  - \alpha } \right)}}{{\Gamma \left( {1 - \gamma  + \beta } \right)\Gamma \left( {1 - \alpha } \right)}}\frac{{\Gamma \left( {2 - \gamma '} \right)\Gamma \left( {\beta ' - \alpha '} \right)}}{{\Gamma \left( {1 - \gamma ' + \beta '} \right)\Gamma \left( {1 - \alpha '} \right)}} \bigg).
\ee
Therefore, the retarded Green's function of the boundary CFT$_4$ is given by
\be
G_R^{\mathrm{AdS_5}}\sim\frac{B(\tilde{\omega}, \tilde{k})}{A(\tilde{\omega}, \tilde{k})}={\left( { - 1} \right)^{ - 2\Delta }}\frac{{\frac{{\Gamma \left( {\alpha  - \beta } \right)}}{{\Gamma \left( {1 - \gamma  + \alpha } \right)\Gamma \left( {1 - \beta } \right)}}\frac{{\Gamma \left( {\gamma '} \right)\Gamma \left( {\beta ' - \alpha '} \right)}}{{\Gamma \left( {\beta '} \right)\Gamma \left( {\gamma ' - \alpha '} \right)}} + \frac{{\Gamma \left( {\beta  - \alpha } \right)}}{{\Gamma \left( {1 - \gamma  + \beta } \right)\Gamma \left( {1 - \alpha } \right)}}\frac{{\Gamma \left( {2 - \gamma '} \right)\Gamma \left( {\beta ' - \alpha '} \right)}}{{\Gamma \left( {1 - \gamma ' + \beta '} \right)\Gamma \left( {1 - \alpha '} \right)}}}}{{\frac{{\Gamma \left( {\alpha  - \beta } \right)}}{{\Gamma \left( {1 - \gamma  + \alpha } \right)\Gamma \left( {1 - \beta } \right)}}\frac{{\Gamma \left( {\gamma '} \right)\Gamma \left( {\alpha ' - \beta '} \right)}}{{\Gamma \left( {\alpha '} \right)\Gamma \left( {\gamma ' - \beta '} \right)}} + \frac{{\Gamma \left( {\beta  - \alpha } \right)}}{{\Gamma \left( {1 - \gamma  + \beta } \right)\Gamma \left( {1 - \alpha } \right)}}\frac{{\Gamma \left( {2 - \gamma '} \right)\Gamma \left( {\alpha ' - \beta '} \right)}}{{\Gamma \left( {1 - \gamma ' + \alpha '} \right)\Gamma \left( {1 - \beta '} \right)}}}},
\ee
which is further simplified into
\be\label{Gr4}
G_R^{\rm{AdS_5}} \sim {\left( { - 1} \right)^{ - 2\Delta }}\frac{{\Gamma \left( { - 2\Delta } \right)}}{{\Gamma \left( {2\Delta } \right)}}\frac{{H\left( {\nu ;\Delta ;\lambda} \right) + H\left( { - \nu ;\Delta ;\lambda} \right)G_R^{\rm{AdS_2}}}}{{H\left( {\nu ; - \Delta ;\lambda} \right) + H\left( { - \nu ; - \Delta ;\lambda} \right)G_R^{\rm{AdS_2}}}}.\ee
Note that  $G_R^{{\rm{Ad}}{{\rm{S}}_{\rm{5}}}} \to {\left( {G_R^{{\rm{Ad}}{{\rm{S}}_{\rm{5}}}}} \right)^{ - 1}}$ as $\Delta \rightarrow -\Delta$.

\subsection{Solution at asymptotic spatial boundary}
The solution of the charged scalar field that approaches the asymptotic spatial boundary of the RN-AdS$_{d+1}$ black hole can be found in the original $r$ coordinate in the region $r_{\mathrm o}\ll L\ll r$, where $f(r)\rightarrow 1$ and $f'(r)\rightarrow 0$. Then, eq.~~(\ref{eomtr}) reduces to
\begin{equation} \label{eomfar}
\phi''(r) +\frac{(d+1)}{r}\phi'(r) = \frac{L^2}{r^2} \left( m^2 - \frac{L^2}{r^2} \vec{k}^2 - \frac{L^2 (\omega + q \mu)^2}{r^2} \right) \phi(r),
\end{equation}
whose solution is given by the Bessel functions as
\be\label{solufar}
\phi(r)&=&\bigg(\frac{L^2\sqrt{\vec{k}^2+(q\mu+\omega)^2}}{r}\bigg)^{\frac{d}{2}}\bigg(\bar{c}_1\Gamma(1-\bar{\nu})
I_{-\bar{\nu}}\bigg(\frac{iL^2\sqrt{\vec{k}^2+(q\mu+\omega)^2}}{r}\bigg)\nno\\
&&+\bar{c}_2\Gamma(1+\bar{\nu})
I_{\bar{\nu}}\bigg(\frac{iL^2\sqrt{\vec{k}^2+(q\mu+\omega)^2}}{r}\bigg)\bigg),
\ee
where $\bar{c}_1$, $\bar{c}_2$ are constants to be fixed by matching the solutions later and $\bar{\nu}\equiv \frac 1 2\sqrt{d^2 + 4 m^2 L^2}$. In the near region ($r\rightarrow\infty$) of the AdS$_{d+1}$ boundary, eq.~(\ref{solufar}) can be expanded as
\be\label{solufar2}
\phi(r)&=&\mathcal{\bar{A}}(\omega, \vec{k})r^{\bar{\Delta}-d}+\mathcal{\bar{B}}(\omega, \vec{k})r^{-\bar{\Delta}},
\ee
where $\bar{\Delta}\equiv\frac{d}{2}+ \bar{\nu}$ is the conformal weight of the boundary operator $\bar{\mathcal{O}}$ dual to the bulk charged scalar field $\phi$, and
\be \label{coeff}
\mathcal{\bar{A}}(\omega, \vec{k})&=&\bar{c}_1\left(\frac{i}{2}\right)^{-\bar{\nu}}
\bigg(L^2\sqrt{\vec{k}^2+(q\mu+\omega)^2}\bigg)^{d-\bar{\Delta}},\nno\\
\mathcal{\bar{B}}(\omega, \vec{k})&=&\bar{c}_2\left(\frac{-i}{2}\right)^{\bar{\nu}}
\bigg(L^2\sqrt{\vec{k}^2+(q\mu+\omega)^2}\bigg)^{\bar{\Delta}}.\ee
In a five-dimensional spacetime ($d=4$), eq.~(\ref{solufar2}) is consistent with eq.~(\ref{phibdy}) in the ``$y$'' coordinate.

\section{CFTs descriptions in IR and UV regions}\label{sec:cft descriptions}
From the AdS/CFT correspondence the IR CFT$_1$ in the near horizon, near extremal limit and the UV CFT$_4$ at the asymptotic boundary of the RN-AdS$_5$ black hole can be connected by the holographic RG flow~\cite{Heemskerk:2010hk,Faulkner:2010jy}. The CFT description of the Schiwnger pair production in the IR region of charged black holes has been systematically studied in a series of previous works~\cite{Chen:2012yn,Chen:2014yfa,Chen:2016caa,Chen:2017mnm}. Here we address the dual CFTs descriptions in the UV region and compare them with those in the IR region.

The IR CFT$_1$ of the RN-AdS black hole is very similar to that of the RN black hole in an asymptotically flat spacetime, since CFT$_1$ can be viewed as a chiral part of CFT$_2$, which has the universal structures for its correlation functions. For instance, the absorption cross section of a scalar operator $\mathcal{O}$ in 2D CFT has the universal form as
\begin{eqnarray} \label{abcft}
\sigma &\sim & \frac{(2 \pi T_L)^{2h_L-1}}{\Gamma(2 h_L)} \frac{(2 \pi T_R)^{2 h_R-1}}{\Gamma(2 h_R)} \sinh\left( \frac{\omega_L - q_L \Omega_L}{2 T_L} + \frac{\omega_R - q_R \Omega_R}{2 T_R} \right)
\nonumber\\
&& \times \left| \Gamma\left( h_L + i \frac{\omega_L - q_L \Omega_L}{2 \pi T_L} \right) \right|^2 \left| \Gamma\left( h_R + i \frac{\omega_R - q_R \Omega_R}{2 \pi T_R} \right) \right|^2,
\end{eqnarray}
where $(h_L, h_R)$, $(\omega_L, \omega_R)$ and $(q_L, q_R)$ are the left- and right-hand conformal weights, excited energies, charges associated with operator $\mathcal{O}$, respectively, while $(T_L, T_R)$ and $(\Omega_L, \Omega_R)$ are temperatures and chemical potentials of the corresponding left- and right-hand sectors of the 2D CFT.
Further identifying the variation of the black hole area entropy with that of the CFT microscopic entropy, namely, $\delta S_{\rm BH} = \delta S_{\rm CFT}$, one derives
\be\label{1stlaw}
\frac{\delta M}{T_H} - \frac{\Omega_H \delta Q}{T_H} = \frac{\omega_L - q_L \Omega_L}{T_L} + \frac{\omega_R - q_R \Omega_R}{T_R},
\ee
where the left hand side of eq.~(\ref{1stlaw}) is calculated in the coordinate (\ref{NRNAdS1}), in which $\delta M=\xi_{\mathrm o}w$, $\delta Q=q$, $T_H=\tilde{T}_{\mathrm n}$ and $\Omega_H=2\mu\ell^2/r_{\mathrm o}$, and thus equals to $w/T_{\mathrm n}-2\pi q_{\mathrm{eff}}\ell $. Moreover, the violation of BF bound in AdS$_2$ makes the conformal weights of the scalar operator $\mathcal{O}$ dual to $\phi$ a complex, which can be chosen as $h_L=h_R=\frac 1 2-i|\nu|$, even without further knowledge about the central charge and $(T_L, T_R)$ of the IR CFT dual to the near extremal RN-AdS$_5$ black hole. One can also see that the absorption cross section ratio (\ref{A}) in the AdS$_2$ spacetime has the form of eq.~(\ref{abcft}) up to some coefficients depending on the mass and charge of the scalar field. With the help of eq.~(\ref{abspair2}), the IR CFT dual of Schwinger pair production $\left| \mathfrak{b} \right|^2$ is the absorption cross ratio of the scalar operator $\mathcal{O}$ with conformal weights changed from $h_{L,R}=\frac12-i|\nu|$ to $h_{L,R}=\frac12 +i|\nu|$~\cite{Chen:2017mnm}.

On the other hand, the absorption cross section and retarded Green's functions in a general 4D finite temperature CFT can not be easily calculated in momentum space as in the 2D CFT. Thus it is not straightforward to compare the calculations from the bulk gravity and the boundary CFT sides. Nevertheless, from eqs.~(\ref{A5}) and (\ref{pair5}), both the absorption cross section ratio $\sigma _{\mathrm{abs}}^{\mathrm{AdS_5}}$ and the Schwinger pair production rate $\left|\mathfrak{b}^{\mathrm{AdS_5}}\right|^2$ calculated from the bulk near extremal RN-AdS$_5$ black hole have simple proportional relation with their counterparts in the near horizon region. Moreover, the violation of BF bound (\ref{BF5}) in the AdS$_5$ spacetime indicates the complex conformal weights $\bar{\Delta}=2+2i|\Delta|$ of the scalar operator $\bar{\mathcal{O}}$ in the UV 4D CFT at the asymptotic spatial boundary of the RN-AdS$_5$ black hole, which also indicates that to have pair production in the full bulk spacetime, the corresponding operators in the UV CFT should be unstable. Interestingly, eq.~(\ref{abspair5}) shows that under the interchange between the roles of source and operator both in the IR and UV CFTs at the same time, namely, $h_{L,R}=\frac12-i|\nu| \to \frac12+i|\nu|$ and $\bar{\Delta}=2+2i|\Delta|\to 2-2i|\Delta|$, the full absorption cross section ratio
$\sigma _{\mathrm{abs}}^{\mathrm{AdS_5}}$ and the Schwinger pair production rate $\left|\mathfrak{b}^{\mathrm{AdS_5}}\right|^2$ interchange with each other only up to a minus sign.

\section{Conclusion}\label{conclusion}
In this paper we have studied the spontaneous scalar pair production in the near extremal RN-AdS$_5$ black hole which possesses an AdS$_2$ structure in the IR region and an AdS$_5$ geometry in the UV region. The solutions for the charged scalar in the full black hole spacetime are obtained by using the matching method, from whose connection the pair production rate, absorption cross section ratio as well as the retarded Green's functions are found at the two boundaries and also for their connection formula. Moreover, the CFTs descriptions of the pair production are investigated both from the AdS$_2$/CFT$_1$ correspondence in the IR and the AdS$_5$/CFT$_4$ duality in the UV regions, and consistent results and new connection between pair production rate and absorption cross section ratio is found, although the related informations computed from the finite temperature 4D CFT is lacking. This work has successfully generalized the study of pair production in charged black holes to the full spacetime and shed a light on a complete understanding to the pair production process in curved spacetime.

\section*{Acknowledgement}
We would like to thank Shu Lin, Rong-Xin Miao and Yuan Sun for useful discussions.
The work of J.R.S. was supported by the NSFC Grant No.~11675272.
The work of C.M.C. was supported by the Ministry of Science and Technology of the R.O.C. under the grant MOST 108-2112-M-008-007.
The work of S.P.K. was supported by National Research Foundation of Korea (NRF) funded by the Ministry of Education (2019R1I1A3A01063183).


\begin{references}

\bibitem{Schwinger:1951nm}
  J.~S.~Schwinger,
  ``On gauge invariance and vacuum polariyation,''
  Phys.\ Rev.\  {\bf 82}, 664 (1951).

\bibitem{Parikh:1999mf}
  M.~K.~Parikh and F.~Wilczek,
  ``Hawking radiation as tunneling,''
  Phys.\ Rev.\ Lett.\  {\bf 85}, 5042 (2000)  [hep-th/9907001].

\bibitem{Chen:2012yn}
  C.-M.~Chen, S.~P.~Kim, I.-C.~Lin, J.-R.~Sun and M.-F.~Wu,
  ``Spontaneous Pair Production in Reissner-Nordstrom Black Holes,''
  Phys.\ Rev.\ D {\bf 85}, 124041 (2012)
  [arXiv:1202.3224 [hep-th]].

\bibitem{Chen:2014yfa}
  C.-M.~Chen, J.-R.~Sun, F.-Y.~Tang and P.-Y.~Tsai,
  ``Spinor particle creation in near extremal Reissner-Nordstrom black holes,''
  Class.\ Quant.\ Grav.\  {\bf 32}, no. 19, 195003 (2015)
  [arXiv:1412.6876 [hep-th]].

\bibitem{Chen:2020mqs}
  C.-M.~Chen and S.~P.~Kim,
  ``Schwinger Effect from Near-extremal Black Holes in (A)dS Space,''
  arXiv:2002.00394 [hep-th].

\bibitem{Chen:2016caa}
  C.-M.~Chen, S.~P.~Kim, J.-R.~Sun and F.-Y.~Tang,
  ``Pair Production in Near Extremal Kerr-Newman Black Holes,''
  Phys.\ Rev.\ D {\bf 95}, no. 4, 044043 (2017)
  [arXiv:1607.02610 [hep-th]].

\bibitem{Chen:2017mnm}
  C.-M.~Chen, S.~P.~Kim, J.-R.~Sun and F.-Y.~Tang,
  ``Pair production of scalar dyons in Kerr-Newman black holes,''
  Phys.\ Lett.\ B {\bf 781}, 129 (2018)
  [arXiv:1705.10629 [hep-th]].

\bibitem{Chen:2009ht}
  C.-M.~Chen, J.-R.~Sun and S.-J.~Zou,
  ``The RN/CFT Correspondence Revisited,''
  JHEP {\bf 1001}, 057 (2010)
  [arXiv:0910.2076 [hep-th]].

\bibitem{Chen:2010bsa}
  C.-M.~Chen, Y.-M.~Huang and S.-J.~Zou,
  ``Holographic Duals of Near-extremal Reissner-Nordstrom Black Holes,''
  JHEP {\bf 1003}, 123 (2010)
  [arXiv:1001.2833 [hep-th]].

\bibitem{Chen:2010as}
  C.-M.~Chen and J.-R.~Sun,
  ``Hidden Conformal Symmetry of the Reissner-Nordstr{\o}m Black Holes,''
  JHEP {\bf 1008}, 034 (2010)
  [arXiv:1004.3963 [hep-th]].

\bibitem{Chen:2010yu}
  C.-M.~Chen, Y.-M.~Huang, J.-R.~Sun, M.-F.~Wu and S.-J.~Zou,
  ``On Holographic Dual of the Dyonic Reissner-Nordstrom Black Hole,''
  Phys.\ Rev.\ D {\bf 82}, 066003 (2010)
  [arXiv:1006.4092 [hep-th]].

\bibitem{Chen:2012ps}
  B.~Chen and J.-J.~Zhang,
  ``RN/CFT Correspondence From Thermodynamics,''
  JHEP {\bf 1301}, 155 (2013)
  [arXiv:1212.1959].

\bibitem{Chen:2012pt}
  B.~Chen and J.-J.~Zhang,
  ``Electromagnetic Duality in Dyonic RN/CFT Correspondence,''
  Phys.\ Rev.\ D {\bf 87}, 081505 (2013)
  [arXiv:1212.1960 [hep-th]].

\bibitem{Chen:2010ywa}
  C.-M.~Chen, Y.-M.~Huang, J.-R.~Sun, M.-F.~Wu and S.-J.~Zou,
  ``Twofold Hidden Conformal Symmetries of the Kerr-Newman Black Hole,''
  Phys.\ Rev.\ D {\bf 82}, 066004 (2010)
  [arXiv:1006.4097 [hep-th]].

\bibitem{Hartman:2008pb}
  T.~Hartman, K.~Murata, T.~Nishioka and A.~Strominger,
  ``CFT Duals for Extreme Black Holes,''
  JHEP {\bf 0904}, 019 (2009)
  [arXiv:0811.4393 [hep-th]].

\bibitem{Hartman:2009nz}
  T.~Hartman, W.~Song and A.~Strominger,
  ``Holographic Derivation of Kerr-Newman Scattering Amplitudes for General Charge and Spin,''
  JHEP {\bf 1003}, 118 (2010)
  [arXiv:0908.3909 [hep-th]].

\bibitem{Garriga:1994bm}
  J.~Garriga,
  ``Pair production by an electric field in (1+1)-dimensional de Sitter space,''
  Phys.\ Rev.\ D {\bf 49}, 6343 (1994).

\bibitem{Pioline:2005pf}
  B.~Pioline and J.~Troost,
  ``Schwinger pair production in AdS(2),''
  JHEP {\bf 0503}, 043 (2005)  [hep-th/0501169].

\bibitem{Kim:2008xv}
  S.~P.~Kim and D.~N.~Page,
  ``Schwinger Pair Production in dS(2) and AdS(2),''
  Phys.\ Rev.\  D {\bf 78}, 103517 (2008)
  [arXiv:0803.2555 [hep-th]].

\bibitem{Cai:2014qba}
  R.-G.~Cai and S.~P.~Kim,
  ``One-Loop Effective Action and Schwinger Effect in (Anti-) de Sitter Space,''
  JHEP {\bf 1409}, 072 (2014)
  [arXiv:1407.4569 [hep-th]].

\bibitem{Kim:2000un}
  S.~P.~Kim and D.~N.~Page,
  ``Schwinger pair production via instantons in a strong electric field,''
  Phys.\ Rev.\  D {\bf 65}, 105002 (2002)
  [arXiv:hep-th/0005078].

\bibitem{Kim:2007pm}
  S.~P.~Kim and D.~N.~Page,
  ``Improved Approximations for Fermion Pair Production in Inhomogeneous Electric Fields,''
  Phys.\ Rev.\ D {\bf 75}, 045013 (2007)
  [hep-th/0701047].

\bibitem{Dumlu:2010ua}
  C.~K.~Dumlu and G.~V.~Dunne,
  ``The Stokes Phenomenon and Schwinger Vacuum Pair Production in Time-Dependent Laser Pulses,''
  Phys.\ Rev.\ Lett.\  {\bf 104}, 250402 (2010)
  [arXiv:1004.2509 [hep-th]].

\bibitem{Kim:2019yts}
  S.~P.~Kim and D.~N.~Page,
 ``Equivalence between the phase-integral and worldline-instanton methods,''
  arXiv:1904.09749 [hep-th].

\bibitem{Balasubramanian:1999}
  V.~Balasubramanian, and P.~Kraus
  ``Spacetime and the Holographic Renormalization Group,''
  Phys.\ Rev.\  Lett {\bf 83}, 3605 (1999)
  [arXiv:hep-th/9903190]


\bibitem{Heemskerk:2010hk}
  I.~Heemskerk and J.~Polchinski,
  ``Holographic and Wilsonian Renormalization Groups,''
  JHEP {\bf 1106}, 031 (2011)
  [arXiv:1010.1264 [hep-th]].

\bibitem{Faulkner:2010jy}
  T.~Faulkner, H.~Liu and M.~Rangamani,
  ``Integrating out geometry: Holographic Wilsonian RG and the membrane paradigm,''
  JHEP {\bf 1108}, 051 (2011)
  [arXiv:1010.4036 [hep-th]].

\bibitem{Faulkner:2009wj}
  T.~Faulkner, H.~Liu, J.~McGreevy and D.~Vegh,
  ``Emergent quantum criticality, Fermi surfaces, and AdS(2),''
  Phys.\ Rev.\ D {\bf 83}, 125002 (2011)
  [arXiv:0907.2694 [hep-th]].

\bibitem{Iqbal:2009fd}
  N.~Iqbal and H.~Liu,
  ``Real-time response in AdS/CFT with application to spinors,''
  Fortsch.\ Phys.\  {\bf 57}, 367 (2009)
  [arXiv:0903.2596 [hep-th]].

\bibitem{Liu:2009dm}
  H.~Liu, J.~McGreevy and D.~Vegh,
  ``Non-Fermi liquids from holography,''
  Phys.\ Rev.\ D {\bf 83}, 065029 (2011)
  [arXiv:0903.2477 [hep-th]].

\bibitem{Faulkner:2011tm}
  T.~Faulkner, N.~Iqbal, H.~Liu, J.~McGreevy and D.~Vegh,
  ``Holographic non-Fermi liquid fixed points,''
  Phil.\  Trans.\  Roy.\  Soc.\ A {\bf  369}, 1640 (2011)
  [arXiv:1101.0597 [hep-th]].

\bibitem{Breitenlohner:1982jf}
  P.~Breitenlohner and D.~Y.~Freedman,
  ``Stability in Gauged Extended Supergravity,''
  Annals Phys.\  {\bf 144}, 249 (1982).

\bibitem{Breitenlohner:1982bm}
  P.~Breitenlohner and D.~Y.~Freedman,
  ``Positive Energy in anti-De Sitter Backgrounds and Gauged Extended Supergravity,''
  Phys.\ Lett.\  B {\bf 115}, 197 (1982).

\bibitem{Xu:1988ju}
  D.~Y.~Xu,
  ``Exact Solutions of Einstein and Einstein-Maxwell Equations in Higher Dimensional Space-time,''
  Class.\ Quant.\ Grav.\  {\bf 5}, 871 (1988).

\bibitem{Hall:1995}
  Leon M. Hall,
  ``Special Functions,''
 https://web.mst.edu/~lmhall/SPFNS/spfns.pdf, page(s):48-49



\end{references}
\end{document}